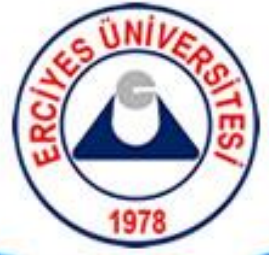

**T.C.**
**ERCİYES ÜNİVERSİTESİ**
**FEN BİLİMLERİ ENSTİTÜSÜ**
**ENERJİ SİSTEMLERİ MÜHENDSİLİĞİ ANABİLİM DALI**

# DÜŞÜK REYNOLDS SAYILI AKIŞLARDA KANAT ÜZERİNDE KARARSIZ AERODİNAMİĞİN İNCELENMESI

**Hazırlayan**
**Gökhan ÖZKAN**

**Danışman**
**Doç. Dr. Mustafa Serdar GENÇ**

**Yüksek Lisans Tezi**

**HAZİRAN 2016**
**KAYSERİ**

**T.C.**
**ERCİYES ÜNİVERSİTESİ**
**FEN BİLİMLERİ ENSTİTÜSÜ**
**ENERJİ SİSTEMLERİ MÜHENDSİLİĞİ ANABİLİM DALI**

# DÜŞÜK REYNOLDS SAYILI AKIŞLARDA KANAT ÜZERİNDE KARARSIZ AERODİNAMİĞİN İNCELENMESİ

**Hazırlayan**
**Gökhan ÖZKAN**

**Danışman**
**Doç. Dr. Mustafa Serdar GENÇ**

**HAZİRAN 2016**
**KAYSERİ**

# TEŞEKKÜR

Tezimin yürütülmesinde ve sonuçlandırmasında yardımlarını ve katkılarını esirgemeyen danışmanım Doç. Dr. Mustafa Serdar GENÇ hocama çok teşekkür ederim.

Erciyes Üniversitesi'ndeki eğitim hayatım boyunca bana destek ve yardımlarını esirgemeyen öğretim üyesi ve elemanlarına teşekkür ederim.

Ayrıca tez çalışmam boyunca desteklerini ve yardımlarını esirgemeyen Rüzgar Mühendisliği ve Aerodinamik Araştırma Ekibi (WEAR) üyelerine teşekkür ederim.

Bununla birlikte Erciyes Üniversitesi Bilimsel Araştırma Projeleri Birimi'ne FYL-2014-5420 kodlu yüksek lisans araştırma projesi ve TÜBİTAK'a 213M329 kodlu 1001 projesi desteklerinden dolayı çok teşekkür ederim.

Eğitim hayatım boyunca maddi ve manevi desteklerini esirgemeyen, bu süre boyunca bana sabırla katlanan ve yardımcı olan aileme sonsuz teşekkürlerimi sunarım.

Gökhan ÖZKAN

Kayseri, Haziran 2016

# DÜŞÜK REYNOLDS SAYILI AKIŞLARDA KANAT ÜZERİNDE KARARSIZ AERODİNAMİĞİNİN İNCELENMESİ

**Gökhan ÖZKAN**

**Erciyes Üniversitesi, Fen Bilimleri Enstitüsü**

**Yüksek Lisans Tezi, Haziran 2016**

**Danışman: Doç. Dr. Mustafa Serdar GENÇ**

## ÖZET

Düşük Reynolds (Re) sayılarında çalışan insansız hava araçları (İHA) ve mikro hava araçları (MHA), rüzgar türbinleri, kuşlar ve böcekler ile ilgili çalışmalar son yıllarda hız kazanmıştır. Bu tez çalışmasında, rüzgar türbini ve mikro hava araçlarında kullanılan NACA4412 profilli en-boy oranı 1 olan (AR=1) kanadın düşük Reynolds sayılı (Re=25000, 50000 ve 75000) akışlarda 0°-45° arasında değişen hücum açıları için kararsız aerodinamiği farklı deneysel çalışmalar yardımıyla incelenmiştir. Kuvvet ölçüm deneyleri sonucunda, kanadın Re=25000, 50000 ve 75000 iken maksimum kaldırma katsayıları sırasıyla 1.19, 1.21, 1.24 ve tutunma kaybı açıları sırasıyla 39°, 38°, 37° olarak bulunmuştur. Duman ve yağ ile akış görselleştirme ve hız dağılımı deneyleri kanat üzerinde farklı düzlemlerde (z/s= +0.4, +0.2, +0.1, 0, -0.1, -0.2) yapılmış. Ayrıca, uç girdapların kanat üzerindeki kararsız akışa etkisi belirlenmiştir. Uç girdapları düşük hücum açılarında kanat uçlarında oluşup iz bölgesine doğru devam ederken hücum açısının artışı ile kanat üzerinde yatay olarak hareket etmekte ve kanat üzerindeki etkisi artmaktadır. Uç girdaplarının kanat gerisinde basınç değişimlerine sebep olması ve kanat üzerinde kanat açıklığı boyunca ilerlemesi ile düşük Re sayılarında kanat üzerinde oluşan ayrılma kabarcıklarını azalttığı ve tutunma kaybını geciktirdiği görülmüştür. Bunların yanında akışın daha kararlı olmasını sağladığı gözlemlenmiştir.

**Anahtar Kelimeler:** Düşük en-boy oranı, uç girdapları, kararsız akış, düşük Re sayısı aerodinamiği, deneysel aerodinamik.

# INVESTIGATION OF UNSTABLE AERODYNAMICS OVER WING AT LOW REYNOLDS NUMBER FLOWS

**Gökhan ÖZKAN**

**Erciyes University, Graduate School of Natural and Applied Sciences**
**M.Sc. Thesis, June 2016**
**Supervisor: Assoc. Prof. Dr. Mustafa Serdar GENÇ**

## ABSTRACT

The number of studies related to unmanned air vehicles (UAV), micro air vehicles (MAV), wind turbines, birds and insects which operate at low Reynolds (Re) numbers have rapidly increased, recently. In this thesis, unsteady aerodynamics of low Reynolds number flows (Re=25000, 50000 and 75000) on NACA4412 airfoil shaped wing which is mainly used in wind turbines and micro air vehicles and operates at low Reynolds numbers and has aspect ratio of 1 (AR=1) was investigated by means of numerous experimental analyses at various angles of attack changing from 0° to 45°. Due to the obtained force measurement results, maximum lift coefficients and stall angles were determined as 1.19, 1.21, 1.24 and 39°, 38°, 37° for Re=25000, 50000 and 75000, respectively. Flow visualization with smoke-wire and oil, velocity distribution experiments were conducted at different planes (z/s= +0.4, +0.2, +0.1, 0, -0.1, -0.2) on the wing. Furthermore, the effects of tip vortices on the wing were determined. The tip vortices were observed at the tip of the wing at low angles of attack and developed through the wake region moved on the wing horizontally with increasing angle of attack and effect of these vortices on the wing increased. It was seen that the tip vortices caused to pressure differences in the wake region and moved to the spanwise direction and this situation led to decrease separation bubble, which is occurred at low Re numbers, and increase stall angle. Additionally, flow condition became steadier due to the tip vortices.

## İÇİNDEKİLER

**DÜŞÜK REYNOLDS SAYILI AKIŞLARDA KANAT ÜZERİNDE KARARSIZ AERODİNAMİĞİN İNCELENMESİ**

### 1. BÖLÜM
### GENEL BİLGİ VE LİTERATÜR TARAMASI

### 2. BÖLÜM
### AMAÇ VE YÖNTEM

# 3. BÖLÜM

## NACA4412 KANAT PROFİLİNİN KARARSIZ AERODİNAMİĞİNİN İNCELENMESİ

# 4. BÖLÜM

## TARTIŞMA – SONUÇ ve ÖNERİLER

# KISALTMALAR VE SİMGELER

| Sembol | Anlamı | Birimi |
|--------|--------|--------|
| Re | Reynolds sayısı | -- |
| St | Strouhal sayısı | -- |
| $C_L$ | Kaldırma kuvvet katsayısı | -- |
| $C_D$ | Sürükleme kuvvet katsayısı | -- |
| c | Kort (veter) boyu | m |
| $V_\infty$ | Serbest akış hızı | m/s |
| S | Kanat alanı | $m^2$ |
| $\rho$ | Hava yoğunluğu | $kg/m^3$ |
| L | Kaldırma kuvveti | $kg.m/s^2$ |
| D | Sürükleme kuvveti | $kg.m/s^2$ |
| N | Normal kuvvet | $kg.m/s^2$ |
| A | Eksenel kuvvet | $kg.m/s^2$ |

# ŞEKİLLER LİSTESİ

# GİRİŞ

Düşük Reynolds (Re) sayılarında çalışan insansız hava araçları (İHA) ve mikro hava araçları (MHA), rüzgar türbinleri, kuşlar ve böcekler ile ilgili çalışmalar son yılarda hız kazanmıştır. Bu araştırmalarda düşük Re sayılarının kullanılmasının nedeni kanatlarının küçük olması veya akış hızlarının düşük olmasıdır. Yüksek irtifalarda olduğu gibi bazı durumlarda ise Re sayısının düşük olmasına hava yoğunluğu ve viskoz etkiler neden olur. Düşük Re sayılı akış durumunda kanat etrafındaki akışta viskoz bölge çok geniştir ve akış içerisinde viskoz kuvvetler baskındır. Kanat üzerine gelen akımı geciktirir ve akış parametrelerindeki küçük değişimler kanat boyunca basınç dağılımında büyük değişimlere neden olur. Düşük Re sayılı akışlarda hücum kenarından firar kenarına doğru ilerledikçe laminer akım viskoz kuvvetlerin etkisi ile çok çabuk bozulmaya başlar. Türbülansa geçiş bölgesinde akım viskoz etkilerin ve ters basınç gradyanlarının üstesinden gelemez ve laminer sınır tabaka ayrılması ve ayrılma kabarcığı meydana gelir. Bu kabarcıkta geri akış ve ters hız profilleri oluşur. Türbülansa geçiş bölgesinde akış gelişip türbülanslı olunca da türbülanslı akışın yeterli enerjisi ile yüzeye tekrar tutunur. Yeniden tutunma ile birlikte akış yüzeye yapışır ve türbülanslı ayrılma olana kadar akışa düzgün bir şekilde devam eder. Ayrılma kabarcığı oluşumu düşük Re sayılarında seyreden hava araçlarında kararsızlığa ve titreşimlere sebep olur ve aerodinamik performansı olumsuz etkiler.

Laminer akıştan türbülanslı akışa geçiş ve ayrılma kabarcığı gibi faktörler, Re sayısının düşük olduğu akışlara maruz kalan araçların performanslarını önemli ölçüde etkiler. Yüksek irtifalarda uçan araçlar yerçekiminin etkisinin azalmasından dolayı aracın ağırlığını taşıması için gerekli olan kaldırma kuvveti azalacaktır. Ancak yüksek irtifalarda havanın yoğunluğunun çok daha fazla azalması ve viskoz etkilerin artmasından dolayı Re sayısı düşecek ve düşük Re sayılarında görülen ayrılma kabarcığı ve tutunma kayıplarının (stall) etkisi ile kanat üzerinde elde edilen kaldırma kuvveti azalırken sürükleme kuvveti artacaktır. Bu etkiler göz önünde bulundurulduğunda araç daha fazla enerji sarfiyatı sağlayacak böylelikle verimliliği azalacaktır.

İHA'lara sensör ve kamera gibi sistemler yerleştirilerek özel amaçlar için kullanılmaktadır. Hatta günümüzde silahlarla dolatılmış İHA'larla karşılaşmamız söz konusudur. Özellikle askeri alanlar için bu sistemler büyük önem taşımaktadır. İHA'nın boyutunun küçük olması ve yüksek irtifalarda uçması İHA'nın düşük Re sayılı akışta uçmasına neden olur. Düşük Re sayısının etkisi ile araç daha fazla enerji tüketecek ve aracın havada kalma süreside azalacaktır. Ancak Düşük Re sayılarında aktif ve pasif akış kontrolü yapılarak ayrılma kabarcığının ve tutunma kaybının etkisi azaltılabilir.

Bu tez çalışması ve 213M329 nolu TÜBİTAK projesi kapsamında en-boy oranı 1 olan NACA4412 profile sahip kanadın kararsız aerodinamiği 25000, 50000 ve 75000 Re sayıları ve 0° – 40° hücum açıları (AoA) için gerçekleştirilmiştir. Yapılan çalışmalarda akış görselleştirme tekniklerinden duman ve yağ deneyleri kullanılmıştır. Ayrıca girdap kopma frekanslarını ve kanat üzerindeki hız dağılımını bulmak için sıcak tel anemometresi ve kuvvet katsayılarının hesaplanması için ise iki eksenli kuvvet ölçüm sisteminden yararlanılmıştır.

# 1.BÖLÜM

## GENEL BİLGİ VE LİTERATÜR TARAMASI

### 1.1.  Giriş

Günümüzde teknolojik gelişmeler ile birlikte havacılık sektörü hızla gelişmiş ve düşük en-boy oranına sahip kanatlar ve plakalar üzerinde akış ve uç girdapları ile ilgili birçok çalışmalar yapılmıştır. Bu çalışmalarda ayrılma kabarcığı, tutunma kaybı ve uç girdaplarının akış performansına etkisi incelenmiştir.

Genç ve arkadaşları [1], NACA2415 kanat profilinin 12°-20° arasında değişen hücum açılarında ve $0.5x10^5$ - $30x10^5$ Reynolds sayıları arasında değişen akış rejimlerinde deneysel çalışmalar yapmışlardır. Bu çalışmada basınç, kuvvet ve hız ölçümleri ve yağ deneyi ile akış görselleştirilmesi yapılmıştır. Çalışma sonucunda hücum açısının artması ile ayrılma kabarcığının ve türbülansa geçişin hücum kenarına doğru kaydığı görülmüştür. Ayrıca Re sayısı artırıldıkça tutunma kaybı karakteristiğinde bir değişme olduğu gözlemlenmiş, düşük Re sayılarında tutunma kaybı açısının yüksek Re sayılarına göre daha düşük olduğu görülmüştür. Düşük Re sayılarında meydana gelen uzun ayrılma kabarcıkları Re sayısı artırıldıkça kısa ayrılma kabarcığı şeklini almıştır.

Katz ve Galdof [2] dikdörtgen NACA-66 hidrofoil üzerinde uç girdap gelişimi ile ilgili bir çalışma yapmıştır. Yapılan bu çalışmada su içerisine enjekte edilen floresan boya lazer yardımı ile yansıtılarak akış görselleştirilmesi gerçekleştirilmiştir. Ayrıca kanat ucu etrafında basınç ölçümleri yapılmıştır. Kanat yüzeyinin pürüzsüz ve pürüzlü olması durumları için deneyler yapmış ve uç girdapların görselleştirilmesinde belirgin bir fark olmamasına karşın basınç farklılığı gözlemlenmiştir.

Uç girdapları etkin hücum açısının azalmasına neden olurlar. Düşük AR ve Re sayılı akışlara maruz kalan kanat profillerinde uç girdapların neden olduğu sürükleme kuvvetleri kanadın aerodinamik performansına büyük oranda etki eder [3]. Viieru ve arkadaşları uç girdaplarının neden olduğu bu sürükleme kuvvetinin etkisinin azaltılması yönünde çalışma yapmışlardır. Uç girdaplarının etkisini azaltmak için kanat ucuna uç plakası konumlandırarak sayısal analizini Re=90000 için gerçekleştirmişlerdir. Yapılan bu çalışma sonucu uç plakasının performansı olumlu yönde etkilediği görülmüştür.

Gim ve Lee [4] NACA0018 kanat profili etrafında meydana gelen uç girdaplarını su tünelinde PIV yöntemi ile hücum açısı 10° ve 20° için ve Re = 2.5 x $10^{4}$'de çalışmalar yapmıştır. Firar kenarına 1C ve 3C uzaklıkta hız profili ölçümleri yapmış ve bu profillerini aynı hücum açıları için kıyaslamalar yapmıştır. Yapılan çalışmada girdap oluşumunun kanat ucundan kanat açıklığının ortalarına kadar ilerlediği gözlemlenmiştir. Kanat ucuna, genişliği maksimum kalınlık kadar olan bir uç plakası takıldığında iz bölgesindeki Reynolds kayma geriliminin azaldığı gözlemlenmiştir. Uç plakasının kanat ortasına kadar olan akış yapısını değiştirdiği görülmüştür.

Liu ve Hsiao [5] AR=1, 2 ve 3 olan dikdörtgen levha üzerinde Re=$10^{4}$ ve $10^{5}$ aralığında kuvvet ölçümü ve akış görselleştirmesi üzerine çalışmalar yapmış (Şekil 1.1) ve uç girdapların ayrılma kabarcığı üzerindeki etkilerini incelemiştir. Yapılan çalışmada uç girdapların laminer ayrılma kabarcığı üzerine bir kuvvet uygulayarak yeniden tutunma sağladığı görülmüştür. Ancak bu kuvvet ayrılma sonrası iz bölgesinin büyük bir kısmını yok ettiği veya değiştirdiği görülmüştür. Yüksek ayrılma açısı ve girdap kaldırması AR=1.6'dan küçük olan levhalarda kaldırma katsayısı eğrisinin doğrusal olmayan bir şekilde artmasına neden olduğu görülmüştür. Yapılan bu çalışmada düşük AR kanatların akış bölgelerinin üç farklı akış özelliğine ayrılabileceği sonucuna ulaşmışlardır. Şekil 1.2.'den görüldüğü gibi hücum açısı 20°'den düşük olduğu durumlarda uç girdaplar kanat ucuna yakın bölgelerde oluşmaktadır. Hücum açısının 20° - 30° arasında olması durumlarında ise Şekil 1.2'de görüldüğü gibi uç girdaplar kanat üzerinde yatay olarak hareket etmektedir. 30°'den yüksek açılarda tutunma devam etmekle birlikte girdap yayılması azalır ve yatay akışın büyük bir kısmının değişimine neden olduğu saptanmıştır. Şekil 1.3'de görülen yağ ile akış görselleştirme deneyinden de anlaşılacağı gibi hücum açısı artırılmaya devam edildiğinde ayrılmanın başladığı ve kanadın büyük bir yüzeyi ayrılmış akış ile kaplandığı görülmüştür. Yapmış olduğumuz tez çalışmasında düz plaka yerine kanat profili kullanılmış ve hız dağılımları ve girdap kopma frekansları da bulunarak akış hakkında ayrıntılı bir inceleme yapılmıştır.

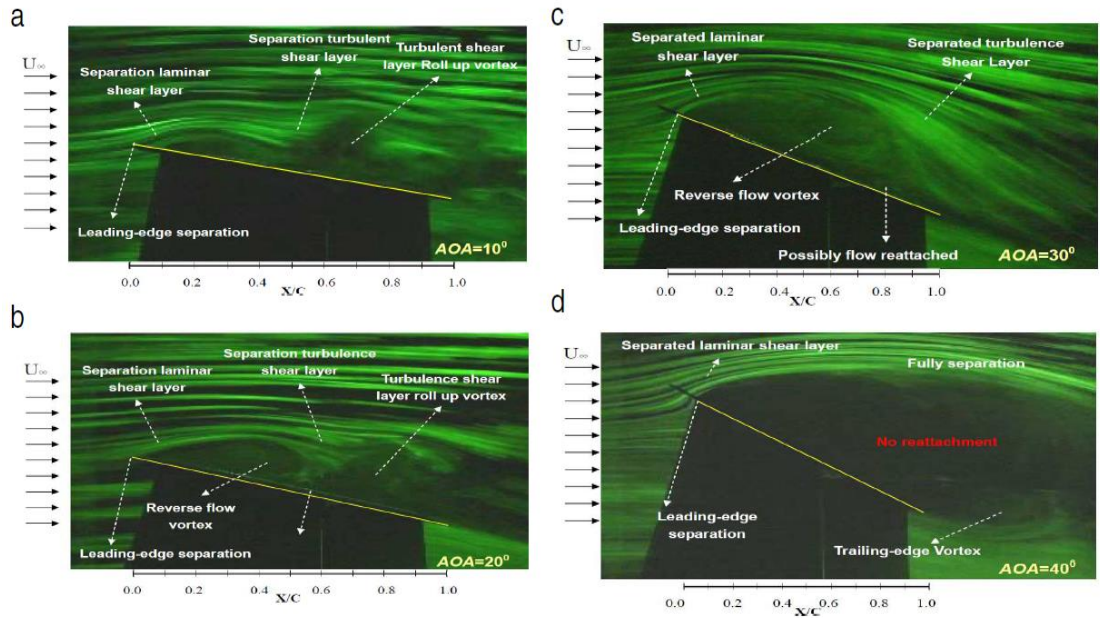

Şekil 1.1. Duman deneyi AR = 1.0 ve Re = $3 \times 10^4$ [5]

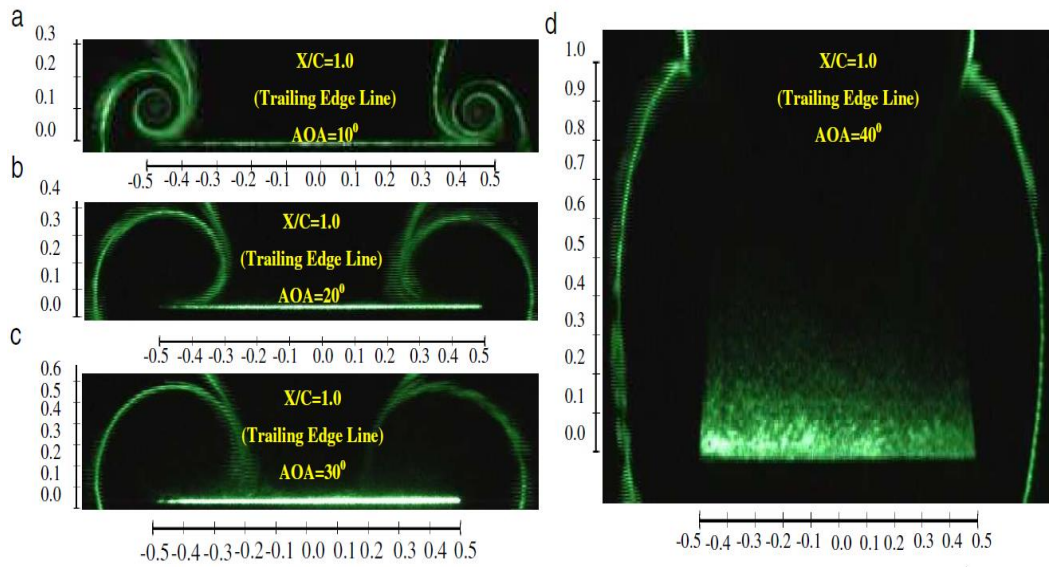

Şekil 1.2. Farklı açılarda duman deneyi firar görüntüsü AR = 1.0 ve Re = $3 \times 10^4$ [5]

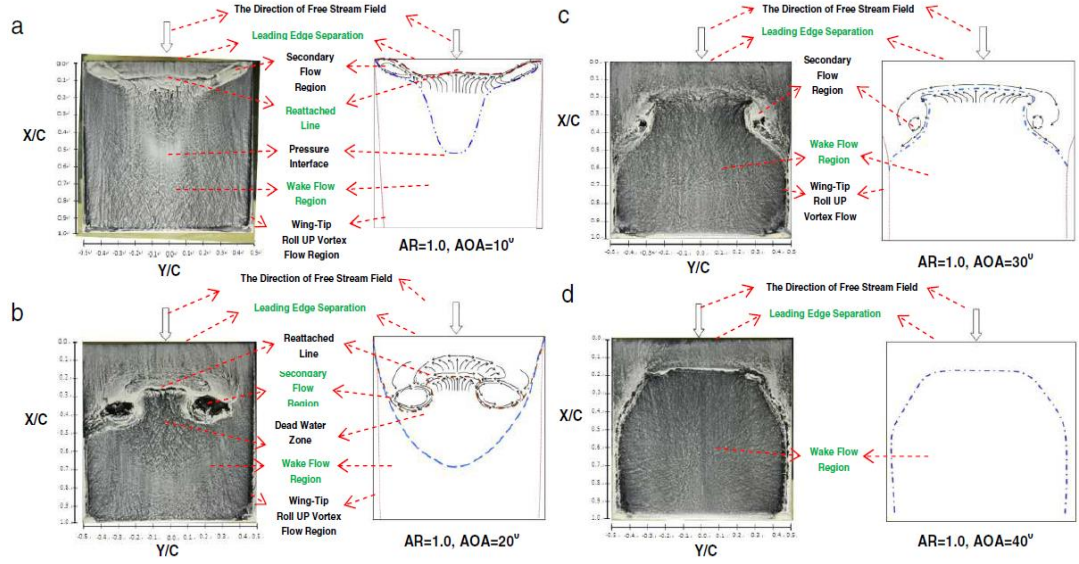

Şekil 1.3. Farklı açılarda kanat üst yüzeyi yağ ile görselleştirme deneyi AR=1.0 ve
Re=1×10⁵ [5]

Bütün MHA'ları için düşük Re sayılı akışlarla ilgili önemli aerodinamik sorunlar vardır. Alam ve Sandham'ın [6] yapmış olduğu çalışmada düz levha üzerinde kararlı (steady) akış ortamında ayrılma kabarcıklarının direk sayısal simülasyonları (Direct Numerical Simulation=DNS) dönen akış oluşturacak kadar kuvvetli kabarcıkların mutlak olarak kararsız olacağını göstermiştir. Türbülans kinetik enerji ve Reynolds gerilme değerleri akışın sadece tekrar tutunmadan sonra üst bölgede oluştuğunu göstermiştir. Ayrılma kabarcığının ters akışın alt katmanlarında simule edilmesine karşın, kararlılık analizi, tam bir kararsız akış için %15 den fazlası ters akış bulunduran bir akış yapısı olduğunu göstermiştir.

Freymuth ve arkadaşları [7] hızlanan ve daimi akışta uç girdapların görselleştirilmesi üzerine çalışmalar yapmışlardır. Deneysel çalışmalarında NACA0015 kanat profili üzerinde duman teli yöntemini kullanmışlardır. Çalışmalarında daimi akış için Re = 5200, hızlanan akışta ise a=2.4 m/s² 'lik bir hızlanma ivmesi kullanmışlardır. Bu çalışmada kanat üzerindeki iki boyutlu ve üç boyutlu girdapları görselleştirmiştir. Hızlanan akışta uç girdap boyutlarının başlangıçta küçük olmasına karşın hızlanma devam ettiğinde giderek büyüdüğü ve bu girdaplar nedeni ile kanadın orta kısımlarında akışın daha kararlı hale getirdiği görülmüştür.

Giuni ve Green [8] NACA0012 kanat profilinin farklı uç şekillerine göre uç girdaplarını incelemek için çalışmalar yapmışlardır. Bu çalışmada akış görselleştirmek için duman teli yöntemi kullanılmıştır. Yapılan çalışmada görülmüştür ki uç girdaplar firar kenarına doğru yayılarak akışı etkilemektedir. İlk girdabın yapısal olarak kendini takip eden ikinci girdaptan daha kararlı olduğu görülmüştür. Kanat ucunun yuvarlatılmış veya düz olması durumlarında aynı denemeler yapılmış ve ucu yuvarlatılmış kanatlarda uç girdapların etkisini azalttığı görülmüştür. Her iki geometri için de farklı hücum açılarında deneyler yapılmış ve hücum açısının artması ile uç girdaplarının etkisinin arttığı görülmüştür.

Kevadiya ve Vaidya [9] yaptığı çalışmada düşük rüzgar hızı olan bölgelerde türbin kanadı seçimi üzerinde durmuşlardır. Rüzgar türbini kanatlarının aerodinamik kanat kesitleri rüzgar türbininin aerodinamik verimliliği üzerinde önemli bir etkiye sahiptir. Bu çalışmada uygun kanat profili olarak NACA4412 tercih edilmiştir. Çalışmada Gambit® ile kanat profili oluşturulmuş ve kaldırma ve sürükleme katsayıları Fluent® yardımı ile $Re=10^5$ için hücum açısı $0°-12°$ arasında hesaplanmıştır. Hesaplamalar sonucu kaldırma/sürükleme oranı $8°$ hücum açısına kadar artarken $8°-12°$ arasındaki hücum açılarında kaldırma/sürükleme oranı azalmaktadır. Düşük hızlar için NACA4412 profilinin $0°-8°$ arasında kullanılması ile uygun bir profil olduğunu belirtmişlerdir.

Mueller [10] mikro hava araçları ile ilgili yaptığı çalışmada MHA'nın küçük yapısından dolayı $Re=2x10^5$'in altında deneyler yapılmıştır. Düz levha ve Eppler 61 kanat profili kullanılarak su ve rüzgar tünelinde deneyler yapılmıştır. Yapılan bu çalışmalar ile iki tünelde de birbirine yakın sonuçlar elde edilmiş ve sistemin her iki tünelde de kararlı bir şekilde çalıştığı gözlemlenmiştir. Kuvvet ve moment ölçümleri yapılarak akış görselleştirme için duman teli ve yağ deneyleri kullanılmıştır. Deneysel çalışmalarda AR=0.5, AR=1, AR=1.5, AR=3 ve sonsuz kanat profili kullanmıştır. Yapılan ölçümler sonucunda $Re = 14×10^4$, $0°-14°$ hücum açıları arasında en yüksek kaldırma kuvvet katsayısı sonsuz kanatta elde edilirken, $14°-22°$ hücum açılarında AR=3 ve $22°-28°$ dereceleri arasında ise AR=0.5 kanatta en yüksek kaldırma katsayısına ulaşmıştır.

Birch ve arkadaşları [11] NACA0015 ve kamburluğu artırılmış kanat profili üzerinde $Re=2.01x10^5$ sayısında deneyler yapmıştır. Deneysel çalışmalarda basınç ölçülmüş ve bu datalar yardımı ile hesaplanan kaldırma ve sürükleme kuvvetleri, kuvvet sistemi ile yapılan ölçümlerle karşılaştırılmıştır. Deneyler sonucunda birincil ve ikincil girdapların birbirleri arasında yüksek bir etkileşim olduğu görülmüştür. Bu etkileşim sonucunda uç bölgesindeki akışın baskın olduğu gözlemlenmiştir. NACA0015 ve kamburlaştırılmış kanat profilinin benzer girdap çaplarına sahip olduğu görülmesine karşın teğetsel hızların NACA0015 kanat profilinde daha düşük olduğu görülmüştür.

Gelişmekte olan, iz bölgesine yakın, üç boyutlu uç girdapları dikdörtgen kanat üzerindeki deneysel analizi üç boyutlu lazer doppler yardımı ile Ramaprian and Zheng [12] tarafından yapılmıştır. Bu çalışmada AR=2 olan NACA0015 kanat profili kullanılmış ve üç bileşenli anlık hız ölçümü yapılarak girdapların akış içerisindeki farklı konum ve kanat açılardaki hız, girdap ve sirkülasyon dağılımları elde edilmiştir. Elde edilen sonuçlar iz bölgesindeki girdapların kayma tabakasındaki oluşumlarını anlamak için kullanılmıştır. Firar kenarına veter boyunun iki katı uzaklıktaki mesafede üç boyutlu girdapların iç kısımlarının yaklaşık olarak eksenel simetrik bir yapı aldığı gözlemlenmiştir. Ayrıca bu çalışmada firar kenarına yakın bölgelerde girdapların oluşumunun başlangıç şartlarına bağlı olarak kaldırma kuvveti ve serbest akım hızının etkisinde olduğu görülmüştür.

Zhi-cheng ve Lei-ping [13] yapmış olduğu çalışmada NACA0015 kanat profili üzerinde $1.8 \times 10^5$ Reynolds sayısında, 8° ve 10° hücum açılarında oluşan uç girdaplarının sayısal analizini sıkıştırılamaz detached eddy simulation (DES) ve Spalart-Allmaras (SA) modeli ile birlikte Reynolds ortalamalı Navier-Stokes (RANS) denklemi sırası ile uygulanarak analizler yapılmıştır. Sayısal çalışma sonuçları Ramaprian ve Zheng [12] yapmış olduğu deneysel çalışmalar ile kıyaslanarak, uç girdap akışı sayısal analizinin doğruluğu kanıtlanmaya çalışılmıştır. Bu çalışma sonucunda DES modelinden elde edilen uç girdaptaki üç boyutlu yapıların, akış içindeki girdapların ve çapraz akış hızlarının deneysel çalışmalarda elde edilen sonuçlarla oldukça iyi bir şekilde örtüştüğü görülmesine karşın RANS-SA modeli ile yapılan analizde sonuçların deney sonuçları ile uyum içinde olmadığı görülmüştür. Yapılan bu çalışma sonucunda DES modelinin uç girdap analizi için uygun bir metot olduğu görülmüş ancak RANS-SA modelinin uç girdap analizi için uygun bir model olmadığı ortaya konulmuştur.

Yarusevych ve Boutilier [14] düşük Reynolds sayılı akışlarda iz bölgesi girdaplarının incelenmesi ilgili bir çalışma yapmıştır. Bu çalışmanın amacı, düşük Reynolds sayısının ve kanat profili kalınlığının girdap kopma frekansına etkisi incelemektir. NACA0018 kanat profili için yapılan bu çalışmada AoA=10° için 30000 ile 200000 Reynolds sayıları arasında deneyler yapılmıştır. Kanat profili iz bölgesinde girdap kopmalarının her iki akış rejimi içinde meydana geldiği ortaya konulmuştur. Girdap kopmasının Reynolds sayısı ile ilişkili olduğu görülmüş, ancak bu uyumun kayma tabakası davranışlarına bağlı olduğu saplanmıştır. Reynolds sayısının artırılması ile akışın kanat üzerinde kayma tabakasında yeniden tutulması sağlanmıştır.

MAV üzerine son araştırmalar farklı derleme makalelerde özetlenmiştir [15], [16]. Aslında, sabit kanat, döner kanat ve çırpan kanat olmak üzere üç çeşit MAV vardır. Bu MAV çeşitleri, değişken serbest akım, hamleler, yunuslama ve dalma (plunging) gibi kararsız durumlarda kanat aerodinamiğini daha iyi anlamaya yardımcı olur. Woods ve diğerlerinin [17] yakın zamanda yapmış oldukları çalışmasında ileri uçuşta çırpan kanat mekanizmasının sabit kanatlı araçlardan daha fazla enerji gerektirdiğini söylemelerine rağmen çırpan kanat mekanizması benzeşimi ile böcek ve kuşların dayanıklılığı ve çevikliğinin araştırılabilmesi mümkün olabilir. Sabit kanat, döner kanat ve çırpan kanat konfigürasyonları için çok düşük Re sayılarında kanat aerodinamiği şimdiye kadar tam anlaşılamamıştır ve özellikle çok düşük Re sayılarında kararsız (unsteady) aerodinamik oldukça ilgi çekmektedir [18], [19]. Rojratsirikul ve diğerleri [20], [21] 2 boyutlu esnek membran kanatların PIV sistemi ile deneysel olarak kararsız aerodinamiğini çalışmışlardır. Bu çalışma sonucunda ortalama membrani şeklin hücum açısına göre fazla değişmediği ve membran kanat titreşim mod ve genliğinin ayrılmış akış tabakasının kararsızlığına ve yerine bağlı olduğu ve membrani kanat salınımı ile akış kararsızlığının etkileştiği gözlemlenmekle birlikte rijit kanat ile esnek kanat karşılaştırmaları sonucunda esnekliğin tutunma kaybını geciktirebileceği sonucuna varılmışlardır. Ayrıca fazla uzunluk oranı olduğunda titreşim modlarının, erken girdap oluşumu ve küçük ayrılmış akış bölgelerinin arttığı görülmüştür.

Kaplan ve arkadaşları [22] düşük AR sahip dikdörtgen, yarı elips ve delta kanat profillerinin firar kenarı girdapları üzerine yapmış oldukları çalışma 8000 ve 24000 Reynolds sayılarında gerçekleştirilmiştir. Bu çalışmada, akış görselleştirmek için boya deneyi ve iz bölgesindeki çapraz akım düzlemlerindeki hız dağılımını ölçmek için PIV sistemi su tünelinde kullanılmıştır. Hız dağılımları kullanılarak kaldırma kuvveti akış ve etkin açıklık vasıtasıyla hesaplanmıştır. Bu hesaplama sonuçları kuvvet sisteminden elde edilen veriler ile kıyaslamışlardır. Böylelikle düşük Reynolds sayılı akışlarda düşük AR sahip kanatların kaldırma kuvveti katsayısının hesaplamalar ile arasındaki uyun gösterilmeye çalışılmıştır. Yapılan çalışmada hesaplamalar ile elde edilen kaldırma kuvvet katsayısının kuvvet sistemi ile elde edilen veriler ile çok iyi bir şekilde örtüştüğü görülmüştür.

Torres ve Mueller [23] yapmış oldukları çalışmada 0.5 - 2.0 arasında 0.25 AR adımlara sahip 4 farklı profil, dikdörtgen, zimmerman, ters zimmerman ve elips, kullanarak $2x10^5$ Reynolds sayısının altında kaldırma kuvvet katsayısı ile ilgili deneyler yapmıştır. Yaptıkları çalışma sonucunda 1.25'AR ye sahip profiller için en yüksek kaldırma kuvvet katsayısı elde edilmesine karşın AR=1 ve daha düşük AR oranlarına sahip kanat profillerinde en yüksek kaldırma kuvvet katsayısına sahip olan profilin ters zimmerman profili olduğu görülmüştür.

AR=0.25 olan düz plaka ve %4 kambur plakanın kaldırma, sürükleme kuvvetlerinin ve yunuslama momentinin ölçümü ile ilgili Pelletier ve Mueller [24] çalışma yapmışlardır. Bu çalışma 60000 - 200000 arasında değişen Re sayılarında gerçekleştirmişlerdir. Yapılan çalışma sonucunda %4 kambur plakanın daha iyi bir aerodinamik karakteristik ve performans gösterdiği görülmüştür.

Şekil 1.4 Farklı en-boy oranına sahip dikdörtgen, zimmerman, ters zimmerman ve elips geometrili kanatlar [23]

## 2. BÖLÜM

## AMAÇ VE YÖNTEM

### 2.1. Amaç ve Kapsam

Bu tez çalışması, deneysel olarak düşük en-boy oranına sahip rijit kanat üzerinde farklı hücum açılarındaki akış karakteristikleri kararsız akış koşullarında incelenmesi üzerine yapılmıştır. Tez çalışmasındaki deneyler 213M329 nolu TÜBİTAK projesinin bir aşaması olarak yapılmıştır. Projenin bu aşamasında en-boy oranı 1 (AR=1) olan NACA4412 kanat profiline sahip kanat kullanılmıştır. Bu profilin tercih edilmesinin sebebi düşük Re sayılarında çalışması ve İHA, MHA ve rüzgar türbinlerinde kullanılmasıdır.

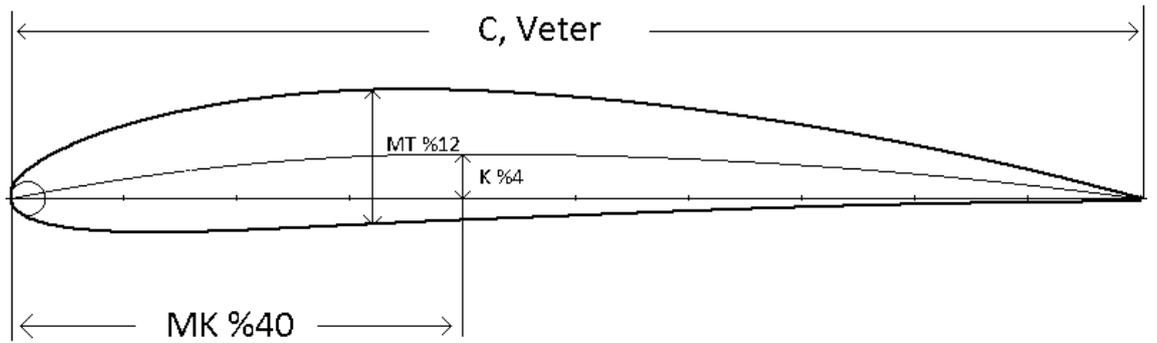

Şekil 2.1. NACA4412 kanat profili

NACA4412 kanat profilin (Şekil 2.1) geometrik özellikler; maksimum kamburluğu veter boyunun %4'ü kadardır, maksimum kamburluk veter uzunluğunun %40'ındadır ve maksimum kalınlık veter boyunun %12'si kadardır. Proje kapsamında kanat üzerindeki akış yağ ve duman teli deneyi yardımıyla görselleştirilmiştir. Kuvvet ölçüm sistemi ile kuvvet katsayıları ölçülerek literatürle karşılaştırılmıştır. Sıcak tel anemometresi ile de kanat ucunda, 1C ve 2C iz bölgesinde ve kanat üzerinde deneyler yapılmıştır.

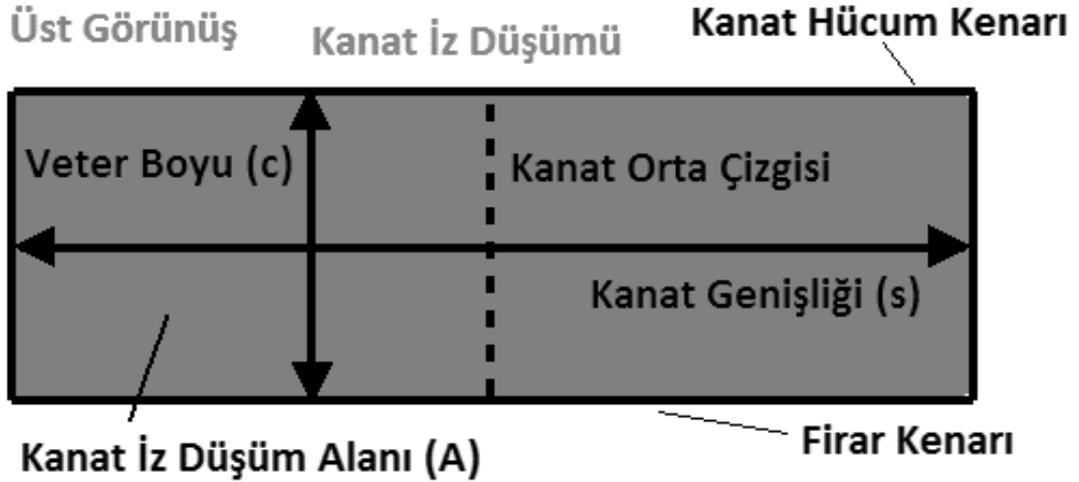

Şekil 2.2. Kanat üzerinde kullanılan tanımlamalar

En-boy oranı hesaplamasında iki önemli faktör vardır, Şekil 2.2'de görüldüğü gibi bunlardan ilki kanat genişliği (s) ve diğeri ise kanat iz düşüm alanıdır (A). En-boy oranı kanat genişliğinin karesinin kanat iz düşüm alanına oranı ile elde edilir ( $s^2/A$ ).

## 2.2. Deneysel Metotlar

### 2.2.1. Kuvvet Katsayısı Ölçümü Deneyi

Havadan yoğun bir cisim uçabilmesi için bir kaldırma kuvvetine ihtiyaç vardır. Bu kaldırma kuvveti uçaklarda, uçak motorundan üretilen itki kuvveti ile uçağın kazandığı hızdan dolayı uçağın yüzeyleri özellikle kanatları üzerindeki hava akımı kaynaklıdır. Uçak hareketli bir şekilde hava içerisinde ilerlerken hava ile uçak arasında bir bağıl hız meydana gelir, kanatların aerodinamik yapısından dolayı kanadın alt yüzeyi ile üst yüzeyi arasında bir basınç farkı oluşur. Oluşan bu basınç farkından dolayı kanat üzerine bir takım kuvvetler etki eder. Bu kuvvetler cismin (uçak, kuş veya böcekler) yerçekimi yönünde olan ağırlığını dengelemek veya cismi dikey yönde hareket ettirmek için oluşan kaldırma kuvveti, cismin hava içerisinde ilerlemesine yardım eden itme kuvvetine zıt yönde olan ve akış nedeniyle oluşan sürükleme kuvvetidir. Kanadın geometrik yapısı değiştirildiğinde etki eden bu kuvvetlerde farklılık gösterir [25].

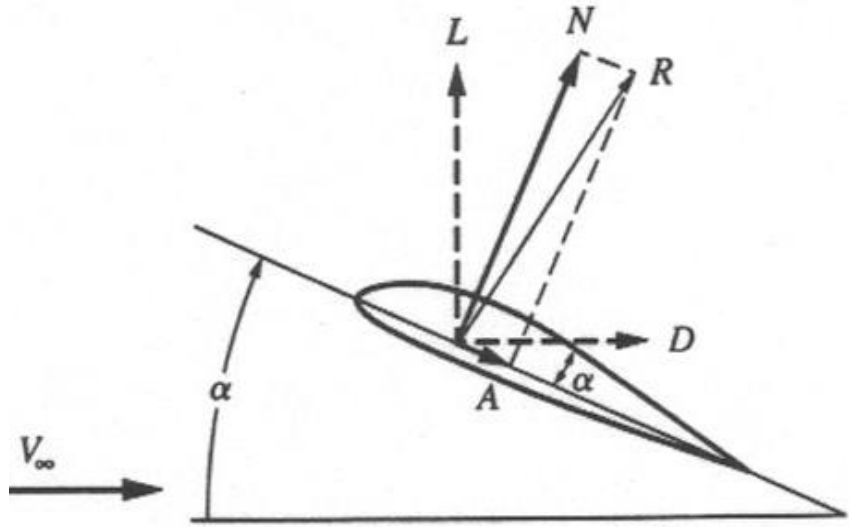

Şekil 2.3. Kanat profili üzerindeki aerodinamik kuvvetler [26]

Kanat üzerinde meydana gelen kuvvetler Şekil 2.3'de görülmektedir. Akış yönü ile kanadın veter düzlemi arasındaki açı hücum açısı (α), akışa dik olan ve kanat profiline etki eden kaldırma kuvveti (L), akışla aynı yönde ve kanada etki eden sürükleme kuvvetidir (D). Bu kuvvetlerin yanında kanat profilinin veter ekseni referans alınarak değerlendirilen kuvvetler de vardır. Kanat profilin kort eksenine dik olan normal kuvveti (N) ve eksene paralel olan eksen kuvvetidir (A) [26].

Kanada etki eden kaldırma ve sürükleme kuvvetleri ile normal ve eksen kuvvetleri arasında vektörel toplam olarak bir eşitlik vardır. Bu vektörel toplama toplam kuvvet (R) adı verilir. Aşağıda verilen denklemler ile bu kuvvetler arasında dönüşüm yapmak mümkündür [26].

$$L = N \cdot \cos(\alpha) - A \cdot \sin(\alpha) \tag{2.1}$$

$$D = N \cdot \sin(\alpha) + A \cdot \cos(\alpha) \tag{2.2}$$

$$N = L \cdot \cos(\alpha) + D \cdot \sin(\alpha) \tag{2.3}$$

$$A = -L \cdot \sin(\alpha) + D \cdot \cos(\alpha) \tag{2.4}$$

Bilimsel çalışmalarda bazı temel denklemler çok karışık olup çoğunun çözümü sayısal yöntemlerle dahi zordur. Karmaşıklığa yol açan terimler akım alanının bazı bölgelerindeki fiziksel büyüklükler korumak şartıyla ihmal edilirse bir takım basitleştirmeler ortaya çıkar ve bu denklemler daha anlaşılır ve kolay hale gelirler [27]. Diferansiyel denklemlerdeki çeşitli terimlerin oransal büyüklüklerini belirlemek için boyutsuz sayılar kullanılır. Bu yüzden aerodinamik kuvvetlerin de herkes tarafından anlaşılır hale gelebilmesi ve boyut karmaşasından kurtulabilmesi için kaldırma ve sürükleme kuvvetini daha genel anlamda ifade eden boyutsuz kaldırma ($C_L$) ve sürükleme ($C_D$) katsayıları kullanılır. Kaldırma ve sürükleme katsayıları aşağıdaki denklemlerle elde edilir.

$$C_L = 2 \cdot \frac{L}{\rho \cdot V_\infty^2 \cdot S} \tag{2.5}$$

$$C_D = 2 \cdot \frac{D}{\rho \cdot V_\infty^2 \cdot S} \tag{2.6}$$

### 2.2.2. Akış Görselleştirme

Akış görselleştirme teknikleri akışın yapısı hakkında görsel olarak analiz yapmamızı sağlamak amacıyla kullanılan yöntemlerdir. Akış görselleştirme tekniklerinden en yaygın olarak kullanılan yöntemler boya, duman teli ve yağ yöntemleridir. Boya deneyi su tünellerinde yapılması kolay olduğundan ve deneylerimizin rüzgar tünelinde yapılmasından dolayı bu tez çalışmasında duman teli ve yağ deneyleri ile akış görselleştirmesi yapılmıştır.

### 2.2.3. Duman Teli Deneyi ile Akış Görselleştirme Deneyi

Duman teli ile akış görselleştirme deneyi rüzgar tünellerinde yaygın olarak kullanılan basit ve verimli bir yöntemdir. Bu deneyde bir duman kaynağı (kanat önüne gerilmiş bir tel yardımıyla elde edilen dumandan veya duman jeneratörü) kullanılarak duman çizgilerinin yani çıkış çizgilerinin hareketi ile akış görsel olarak karakterize edilir. Bu deneyde dikkat edilmesi gereken en önemli husus duman kaynağının akışı bozmayacak yapıda olması gerekmektedir. Duman telinin akışa engel oluşturacak kalınlıkta olması akışımızın türbülanslı olmasına neden olacaktır.

### 2.2.4. Yağ Deneyi ile Akış Görselleştirme Deneyi

Yağ ile akış görselleştirme yöntemi bize akışın kanat yüzeylerinden geçerken oluşturduğu etkiyi gösteren basit bir yöntemdir. Gazyağı içerisine katılan renklendiriciler ve karışımın homojen olmasını sağlayan yardımcı maddeler katılarak düşük viskoziteli bir karşım elde edilir. Akışkan içerisinde bulunan geometrik şekil üzerine akış yok iken bu düşük viskoziteli karışım sürülerek, akış ile akışkanın yüzey üzerindeki etkileri kolay bir şekilde görselleştirilebilir. Ancak bu yöntemin de bazı dezavantajları vardır, örneğin çok düşük hızlı akışlarda karışımın viskozitesi nedeni ile akışkanın etkisi tam olarak görülemeyebilir.

### 2.2.5. Sıcak Tel Anemometresi ile Hız Ölçümü (Hotwire) Deneyi

Sıcak tel anemometresi, 5µm kalınlığındaki tungsten tel üzerinde meydana gelen ısı transferi hızı ile akışkanın hızını ilişkilendirerek hızın anlık olarak ölçülmesini sağlayan deney sistemidir. Bu sistemlerde farklı ortamlarda (sıvı veya gaz içerisinde) ölçüm yapmamızı sağlayan özel probları vardır ve bu problar yardımı ile sistem özelliğine bağlı olarak saniyede 250000 örnekleme ile ölçüm yapmamıza olanak tanımaktadır. Sistemin çalışma prensibi prob ucundaki tellere verilen elektrik akımı ile tel ısıtılır ve daha düşük sıcaklıktaki akışkan tel üzerinden geçerek zorlanmış taşınıma neden olur. Sıcaklığı düşen telin probun bağlı olduğu sistem sabit tutmaya çalışır ve sistemde bulunan wheatstone köprüsü yardımı ile prob üzerinden geçen akım gerilim sinyaline çevrilerek bilgisayarda bağlı bulunan analog dijital çeviriciye gönderilir. Bu çevirici gönderilen gerilim sinyallerini bilgisayar dili olan ikili sayı sistemine çevirir. Bilgisayar yazılımı yardımıyla yapılan kalibrasyon probtan gelen bu sinyallerini hız değerlerine dönüştürür [28].

Sıcak tel anemometresi deneyinde çok fazla veri çok kısa bir sürede ölçülebildiği için deney sisteminde meydana gelen girdaplar ve akış içindeki frekans değişimleri bu sistem yardımı ile ölçülebilmektedir.

## 3. BÖLÜM

## NACA4412 KANAT PROFİLİNİN KARARSIZ AERODİNAMİĞİNİN İNCELENMESİ

### 3.1. Giriş

Bu tez çalışmasında NACA4412 profile sahip kanadın kararsız akış karakteristiği ve uç girdapları ile ilgili yapılan deneysel yöntemler bahsedilmiştir ve bu girdapların etkisi incelenmiştir.

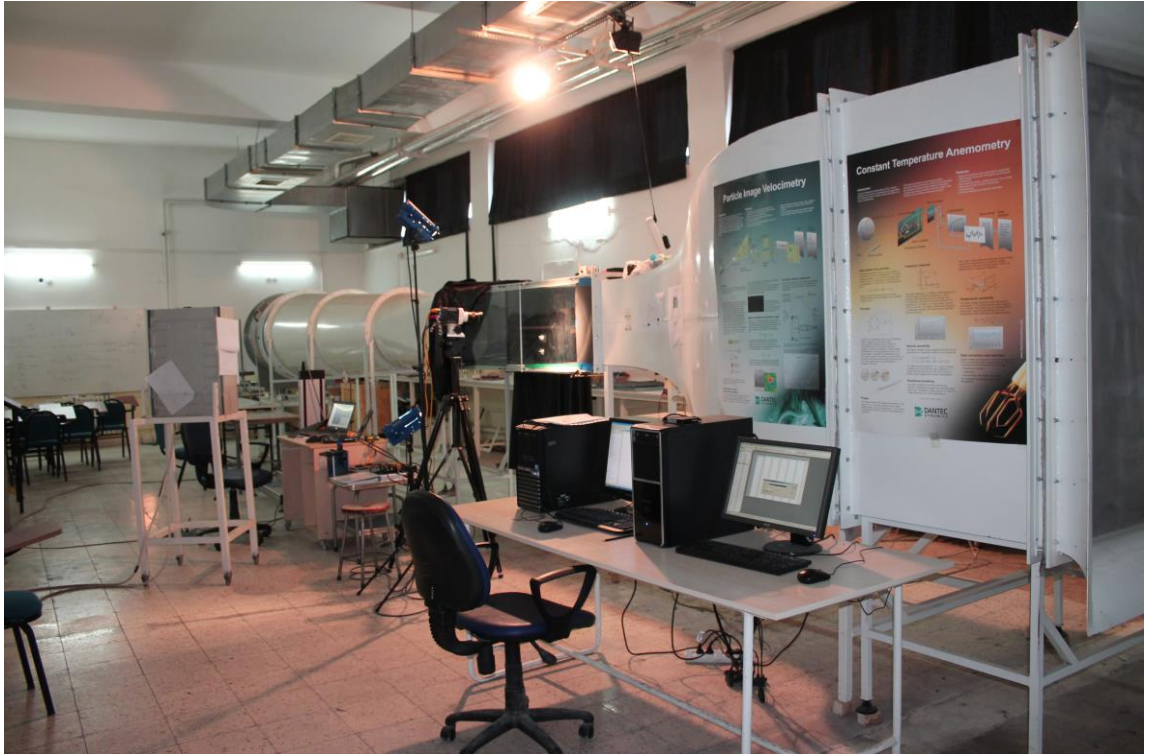

Şekil 3.1. Deneylerde kullanılan rüzgar tüneli.

Tez çalışmasındaki deneyler Erciyes Üniversitesi Mühendislik Fakültesi Enerji Sistemleri Mühendisliği Bölümü Rüzgar Mühendisliği ve Aerodinamik Laboratuvarında mevcut olan emmeli tip açık çevrim rüzgar tünelinde (Şekil 3.1) yapılmıştır. Rüzgar tünelinin özellikleri şu şekildedir; 200 cm tünel test odası uzunluğu, 50 cm x 50 cm şeffaf test odası, tünel fan gücü 15 kW, devri 1500 d/d ve hız ayarı bir kontrol sistemi ile yapılabilmektedir.

## 3.2. Deneysel Kontroller

### 3.2.1. Rüzgar Tüneli Test Odası Karakteristikleri

Test odasının hız dağılımı ve türbülans seviyesi Karasu'nun [28] yapmış olduğu tez çalışmasında ölçüldüğü üzere türbülans şiddeti Şekil 3.2'de görüldüğü gibi test bölgesinde 5m/s hızlar için %0.7 çıkmasına karşın daha yüksek hızlarda %0.4 oranında ölçülmüştür. Bu değerlere göre tünelin türbülans seviyesinin istenilen seviye içerisinde olduğu görülmüştür.

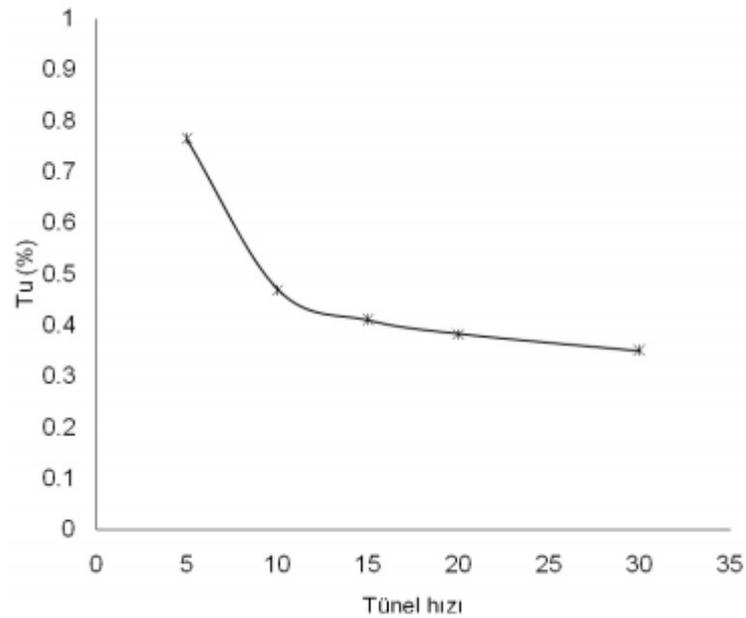

Şekil 3.2. Rüzgar tüneli test odası merkezindeki türbülans dağılımı [28].

### 3.2.2. Belirsizlik Analizi

Deneysel sonuçların geçerliliğinin belirlenebilmesi için mutlaka bir hata analizi yapmak gerekmektedir. Deneylerden elde edilen veriler kullanılarak hesaplanan parametrelere ait hata oranlarının tespiti için pratikte bir kaç yöntem geliştirilmiştir. Bunlardan en yaygın olarak kullanılan belirsizlik analizi yöntemidir. Bu yöntemde Denklem 3.1 kullanılarak maksimum hata oranı belirlenir.

$$u_R = \left[ \left( \frac{x_1}{R} \frac{\partial R}{\partial x_1} u_1 \right)^2 + \left( \frac{x_2}{R} \frac{\partial R}{\partial x_2} u_2 \right)^2 + \cdots + \left( \frac{x_n}{R} \frac{\partial R}{\partial x_n} u_n \right)^2 \right]^{0.5} \tag{3.1}$$

Bu çalışmada hız ve kuvvet ölçümü yapılmıştır ve bu bileşenlere bağlı olan kuvvet katsayısı, Re sayısı için belirsizlik analizi yapılmıştır. Yapılan çalışmada kullanılan 25000, 50000 ve 75000 Re sayıları için belirsizlik analizi yapılarak hata payları yüzde cinsinden bulunmuştur. Re sayıları için belirsizlik analizi sonuçları sırası ile 5.9, 5.5, 5.2, $C_L$ için 6.0, 5.7, 5.4 ve $C_D$ için 6.7, 6.2, 5.8 bulunmuştur.

### 3.2.3.  NACA4412 Profile Sahip Kanat Üretimi

Kanat üretimi için Erciyes Üniversitesi Mühendislik Fakültesi Enerji Sistemleri Mühendisliği Bölümü Rüzgar Mühendisliği ve Aerodinamik Laboratuvarında bulunan 3D yazıcı (Şekil 3.3) ile kanat üretilmiştir.

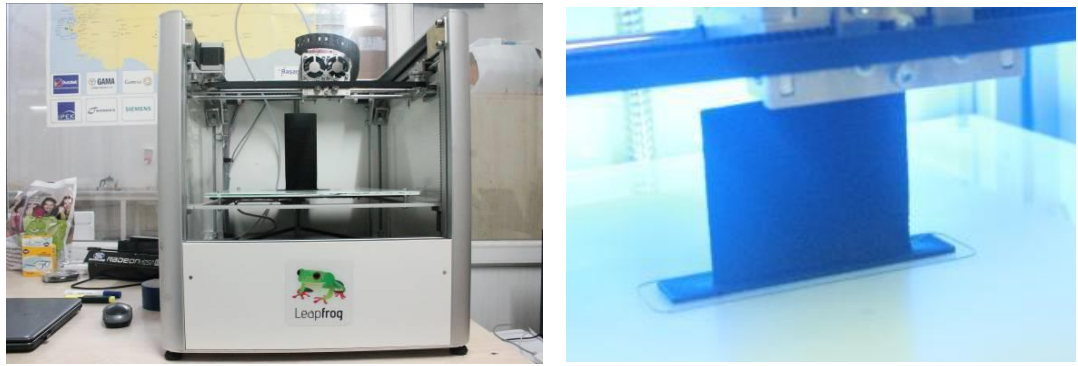

Şekil 3.3. 3D yazıcı ile NACA4412 kanat üretimi

Kanat üretiminde dikkat edilen hususlardan biri termal gerilmelerden dolayı kanat boyutunda meydana gelecek sapmalardı. Bu yüzden kanat üretimi birkaç deneme sonucunda yapılmıştır ve kanat ölçüleri standartlara uygun bir şekilde üretilmiştir. Kanat üretildikten sonra kanadın pürüzlü yüzeyleri zımpara yardımı ile düzletilerek boyanmıştır. Hazırlanan kanat tünele bağlanarak deneysel ölçümler için hazırlanmıştır (Şekil 3.4).

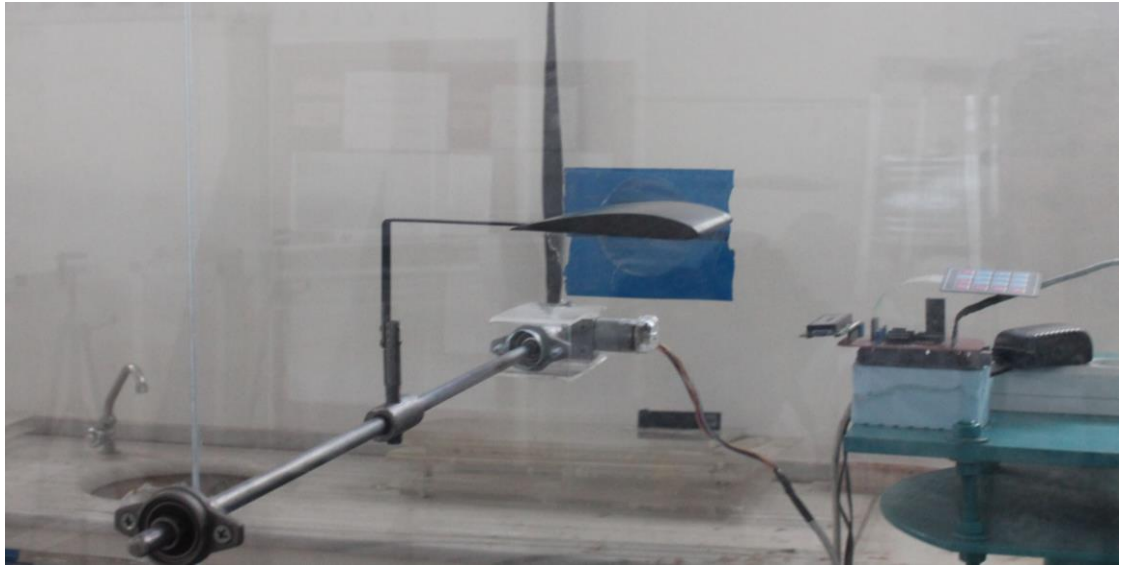

Şekil 3.4. Test odasına deney için bağlanmış NACA4412 profile sahip kanat

### 3.2.4. Aerodinamik Kuvvetlerin Ölçümü

Kanadın AR=1 için farklı Reynolds sayılarında (Re=25000, 50000 ve 75000) ve 0°'den 45°'ye kadar belirli hücum açılarında aerodinamik kuvvetleri kendi tasarımımız olan kuvvet sistemi ile (Şekil 3.5) ölçülmüştür. Bu sistemde iki adet strain gauge eksenel (veter hattına dik N ve paralel A) kuvvetleri ölçmektedir. Sistem 200 Newton'a kadar olan kuvvetleri ölçebilmekte ve saniyede 100 ölçüm alarak 10 saniyelik ölçüm ortalamasını vermektedir. Ölçülen kuvvetler daha sonra x ve y eksenine (Denklem 2.1 ve 2.2) transfer edilmiş ve $C_L$, $C_D$ (Denklem 2.5 ve 2.6) katsayıları hesaplanmıştır.

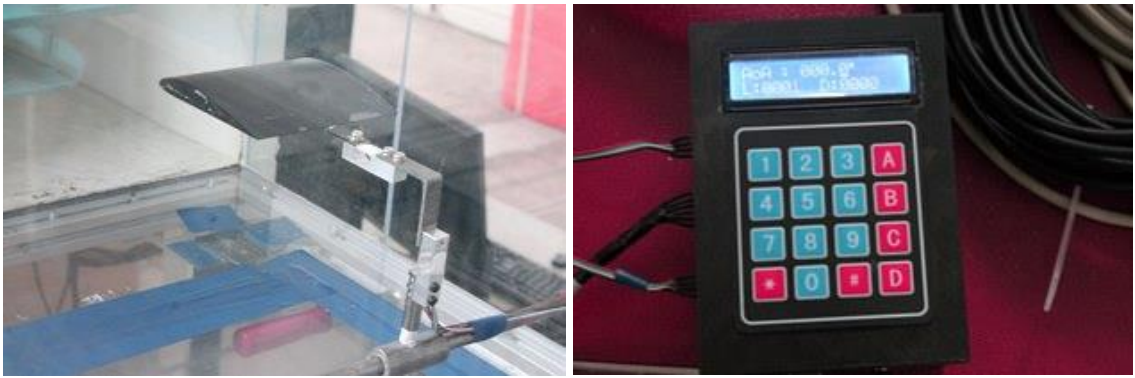

Şekil 3.5. Kuvvet ölçüm sistemi ve elektronik göstergesi

Şekil 3.6a'da NACA4412 kanadın farklı Re sayılarında $C_L$ ve $C_D$ katsayılarının hücum açısı ile değişim grafikleri verilmiştir. Bu grafiklerde elde ettiğimiz sonuçlara göre Re=25000, 50000 ve 75000 iken tutunma kaybı açıları Şekil 3.6a'da görüldüğü gibi sırasıyla 39°, 38° ve 37°'dir ve Re sayısının artışı ile $C_{Lmaks}$ değerleri 1.19, 1.21, 1.24 olarak belirlenmiştir. Bu sonuçlardan; bilinen Re sayısı artıkça $C_{Lmaks}$ değerlerinin arttığı sonucu ortaya çıkmaktadır. Aynı zamanda her bir Re sayısında farklı olacak şekilde 15°-20° hücum açısı aralığında kanat kaldırma kuvveti performansında bir düşüş, sürükleme kuvvetinde ise ani artma görülmektedir. Bu performans değişikliğinin düşük Re sayılarında görülen laminer ayrılma kabarcığının büyümesi olabileceği düşünülmektedir.

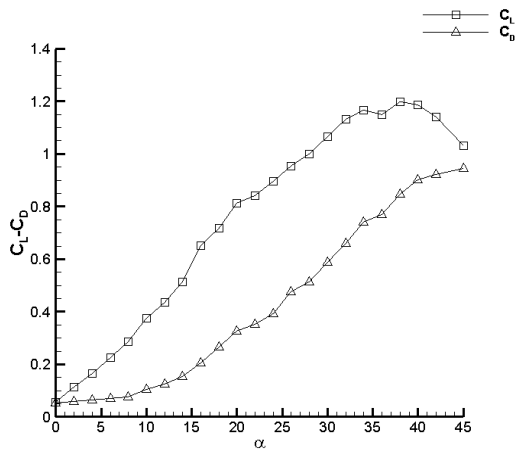

Re = 25000

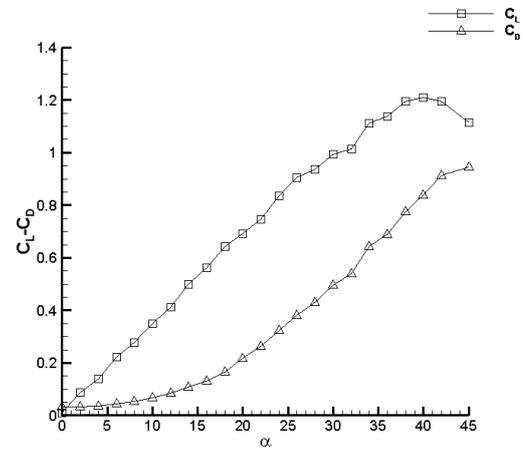

Re = 50000

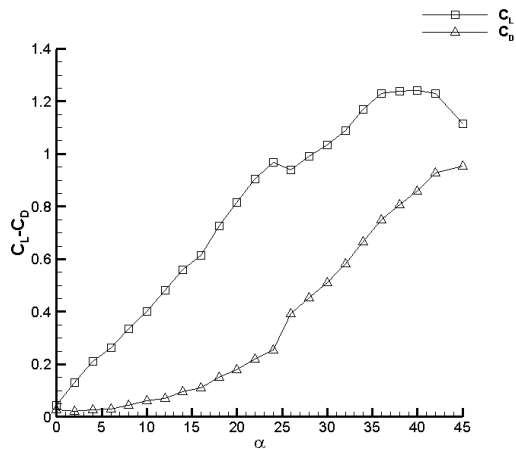

Re = 75000

Şekil 3.6a. Farklı Re sayılarında aerodinamik kuvvet katsayıları

Şekil 3.6b'de görüldüğü gibi kanadın farklı Re sayılarındaki L/D oranı görülmektedir. Re sayısı arttıkça kanadın kaldırma ve sürükleme kuvvet katsayılarının oranının arttığı görülmüştür. Re sayıları 25000, 50000 ve 75000 için maksimum L/D oranları sırasıyla 8°, 8° ve 7° hücum açılarında 3.95, 5.32 ve 9.33 olarak hesaplanmıştır. Bu durumda kanadın 8° de kullanılması uygun görülmektedir.

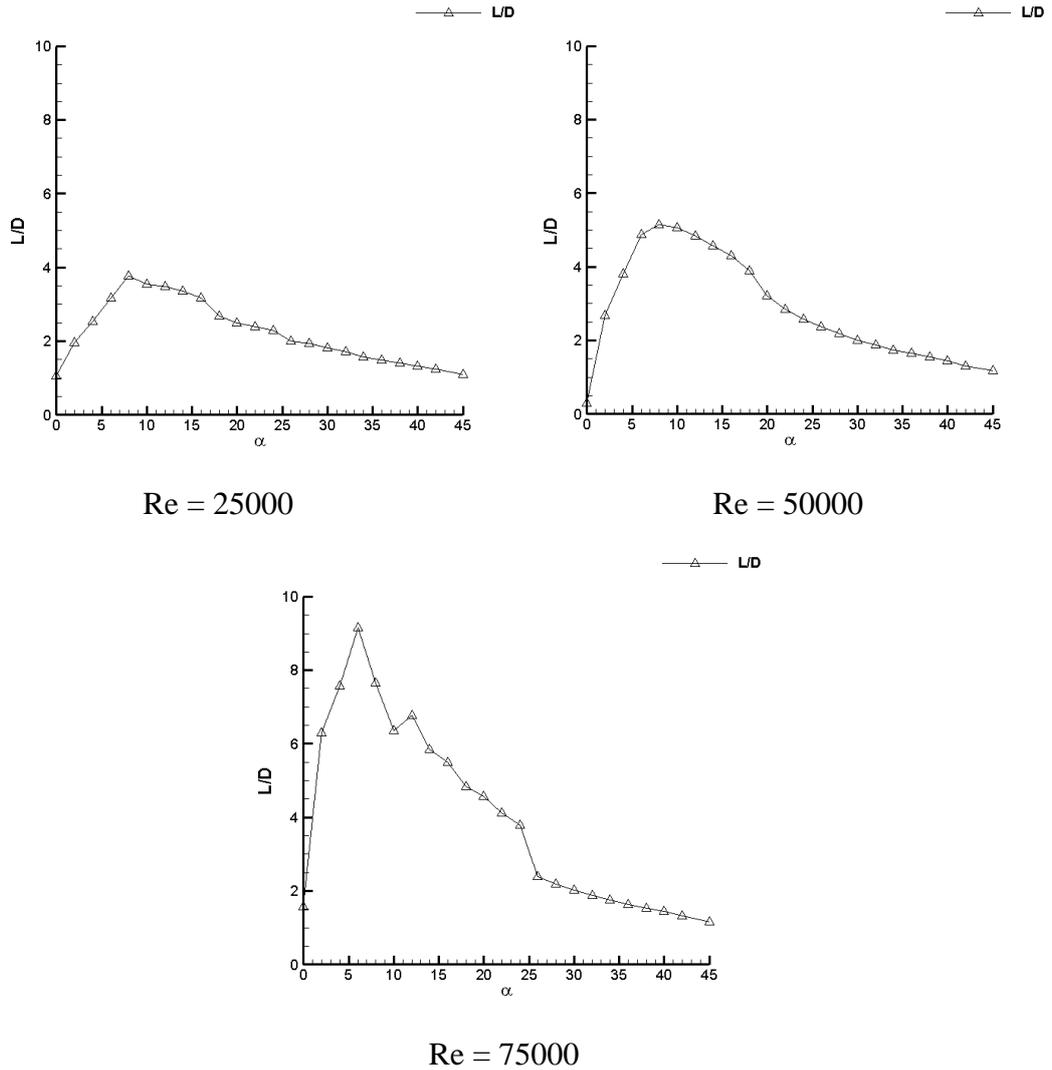

Re = 25000        Re = 50000

Re = 75000

Şekil 3.6b. Farklı Re sayılarında L/D oranı

Hücum açısı arttıkça uç girdapların etkisi kanat yüzeyine yayılmakta ve kanadın firar kenarının büyük bir kısmında etkili olmaktadır. Bu girdaplardan dolayı meydana gelen basınç düşüşü firar bölgesinde vakum etkisi yapmakta ve $C_L$ ve $C_D$ grafiklerinden de görüldüğü gibi sürükleme kuvveti artmaktadır.

### 3.2.5. Kanat Üzeri Hız Dağılımı

Kanat üzerinde hız dağılımı ölçümleri için parçacık görüntülemeli akış ölçüm cihazı (PIV) yaygın olarak kullanılmaktadır. Ancak bu sistemin pahalı olması ve şu an için Erciyes Üniversitesi Mühendislik Fakültesi Enerji Sistemleri Mühendisliği Bölümü bünyesinde bulunmamasından dolayı, bölümümüzün Rüzgar Mühendisliği ve Aerodinamik Laboratuvarında bulunan sıcak tel anemometresi travers sistemi ile kullanılmıştır. Sıcak tel probu olarak 0-20 m/s hız aralığında ölçüm yapabilen rüzgar tüneli için uygun Dantec® marka 55P51 kodlu çift eksenli prob kullanılmıştır. Çift eksenli problar yardımı ile akışın iki eksene ait hızları ölçülebilmektedir.

Hız dağılımı deneyleri kanat üzerindeki hız dağılımını, uç girdaplarının kanat üzerindeki akışa etkisini örnek olarak görebilmek için sadece 8° hücum açısı için yapılmıştır. Kanat üzerinde aslında 3 boyutlu akış durumu (en boyunca farklı noktalarda alan taraması ile) çıkartılması hedeflenmiştir. Bunun nedeni uç girdaplarının kanat üzerindeki ayrılma kabarcığına etkisi araştırılmasıdır. Konum belirlenirken; kanat simetri ekseninde yani z/s=0.5'de kanat tutturma aparatı olduğu için z/s=0.4 kanadın ortası için ele alınmış, z/s=0.1 kanat ucundan biraz içeride kanat üstündeki bölge, z/s=0.0 kanadın tam uç kısmı ve z/s=-0.1 ise kanadın dışı olarak ele alınmıştır. Şekil 3.7 ve 3.8'de ölçüm yapılan alanlar görülmektedir.

Sıcak tel probu kanat yüzeyine yaklaştırılabildiği kadar yaklaştırılmış (kanat yüzeyinden 1.5 mm yukarıya) ve anemometre ile hızın x ve y yönündeki hız bileşenleri dikey eksende hareket ederek ölçülmüş ve türbülans yoğunluğu ve Reynolds gerilmeleri hesaplanmıştır. Daha sonra bütün bu elde edilen sonuçlar alınan nokta değerlerinden konturlara dönüştürülmüştür.

Şekil 3.7 ve 3.8'deki grafiklerde, kanadın değişik Re sayılarında hız dağılımları yani yatay hız vektörlerinden elde edilen hız konturları görülmektedir. Bu hız dağılımlarına baktığımızda, uç girdapları dolayısıyla kanadın ucuna yakın z/s=0.1 ve z/s=0.0 konumlarındaki hız alanlarında özellikle düşük Re sayılarında hız artışı kanat gerisinde dikkat çekmektedir.

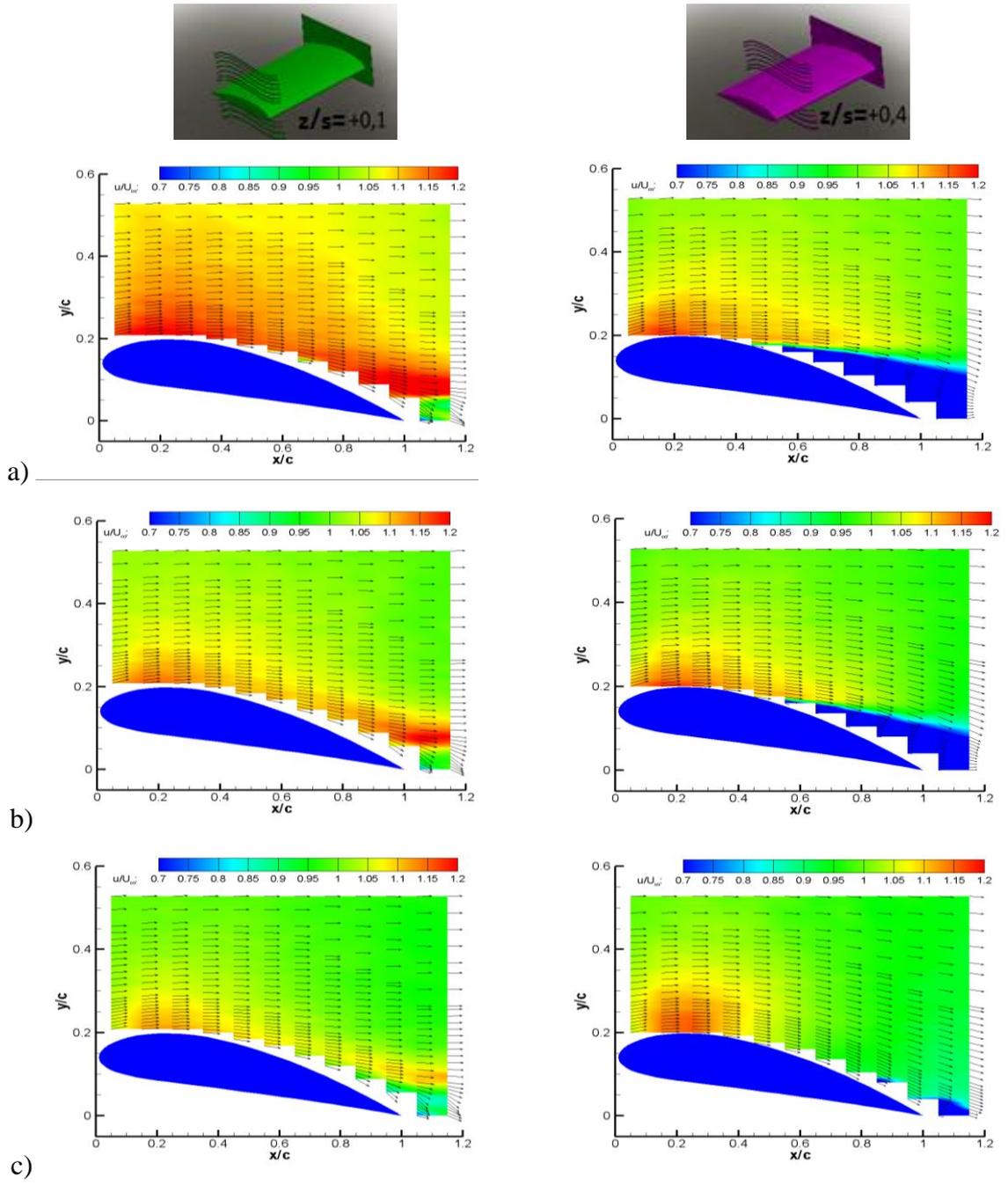

Şekil 3.7. Kanat üzerinde farklı düzlemlerde hız dağılımı α=8° (a) Re=25000, b) 50000
c) 75000

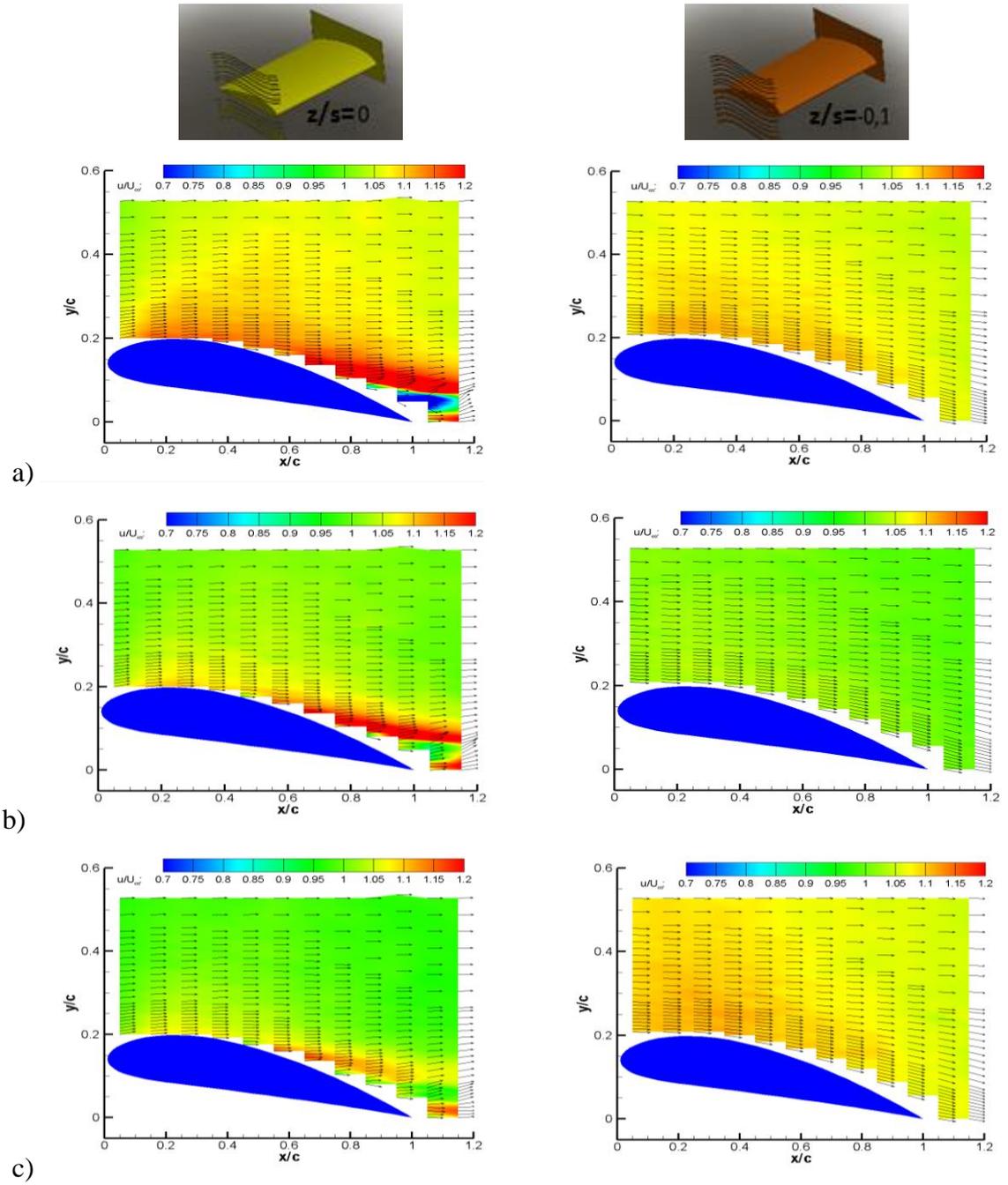

Şekil 3.8. Kanat üzerinde farklı düzlemlerde hız dağılımı α=8° (a) Re=25000, b) 50000 c) 75000

Re = 50000 ve 75000 için atalet kuvvetleri 25000 Re sayısındaki atalet kuvvetlerine göre fazla olduğundan oluşan uç girdapları yatay hareket edemeden kanat gerisine gitmektedir. Kanadın z/s=0.4 konumundaki hız dağılımlarında ise firar kenarında daha yakın oluşmuş ayrılma kabarcıkları ve bu kabarcık bölgesindeki hız azalmaları görülmektedir. Ayrıca kabarcık ve hız azalma bölgeleri Re sayısı arttıkça küçülmekte ve kabarcık yüksekliği azalmaktadır. Kanat üzerinde hız dağılımları kanadın hemen üzerinde her durum için standart olarak alınamadığı için mavi renkli bölgelere bakılınca yanıltıcı olabilmektedir.

Farklı Re sayıları için elde edilen hız dağılımlarından hesaplanmış olan türbülans yoğunluğu ve Re gerilmelerine ait grafikler Şekil 3.9a, 3.9b ve 3.9c'de verilmiştir. Bu grafiklerden görüldüğü gibi; z/s=0.4 düzleminde yani kanadın ortasına yakın bölgede Re sayısı arttıkça atalet kuvvetinin artışı ile kanat üzerinde girdap etkilerinin azalmasından dolayı türbülans yoğunluğu ve Re gerilmelerinin azaldığı görülmektedir. Bunun aksine z/s=0.1 ve z/s=0 düzlemlerinde yani kanadın ucuna yakın ve tam ucundaki düzlemlerde uç girdaplarının etkisi ile kanat firar kenarında türbülans ve Re gerilmesi değerlerinin Re sayısı artışı ile arttığı görülmüştür. Bu artışın kanat uçlarındaki uç girdapları dolayısıyla 3 boyutlu akış kaynaklı olduğu açıkça görülmektedir.

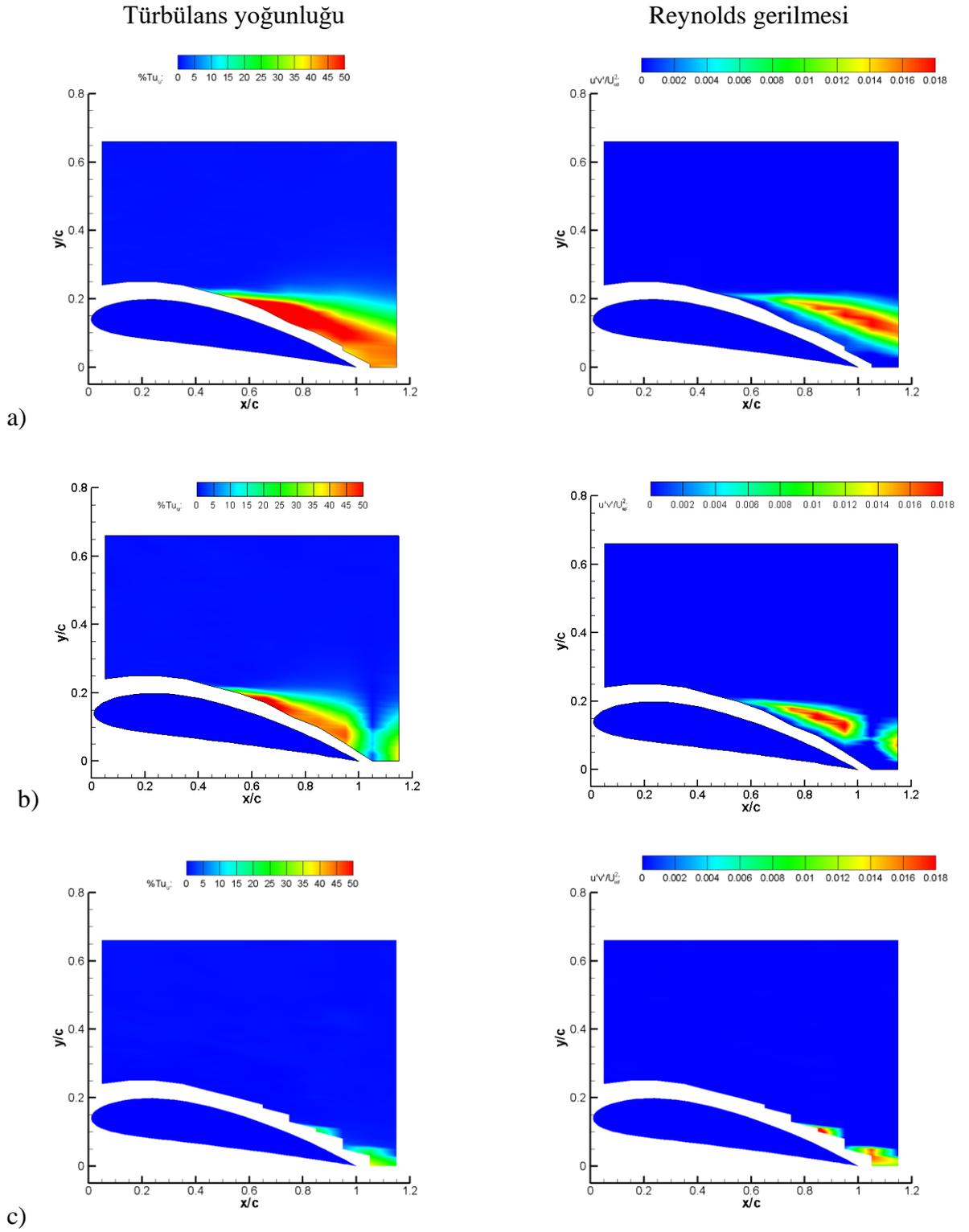

Şekil 3.9a. z/s=0.4'de kanat üzerinde türbülans yoğunluğu ve Reynolds gerilmesi
dağılımları α=8° a) Re=25000 c) 50000 b) Re=75000

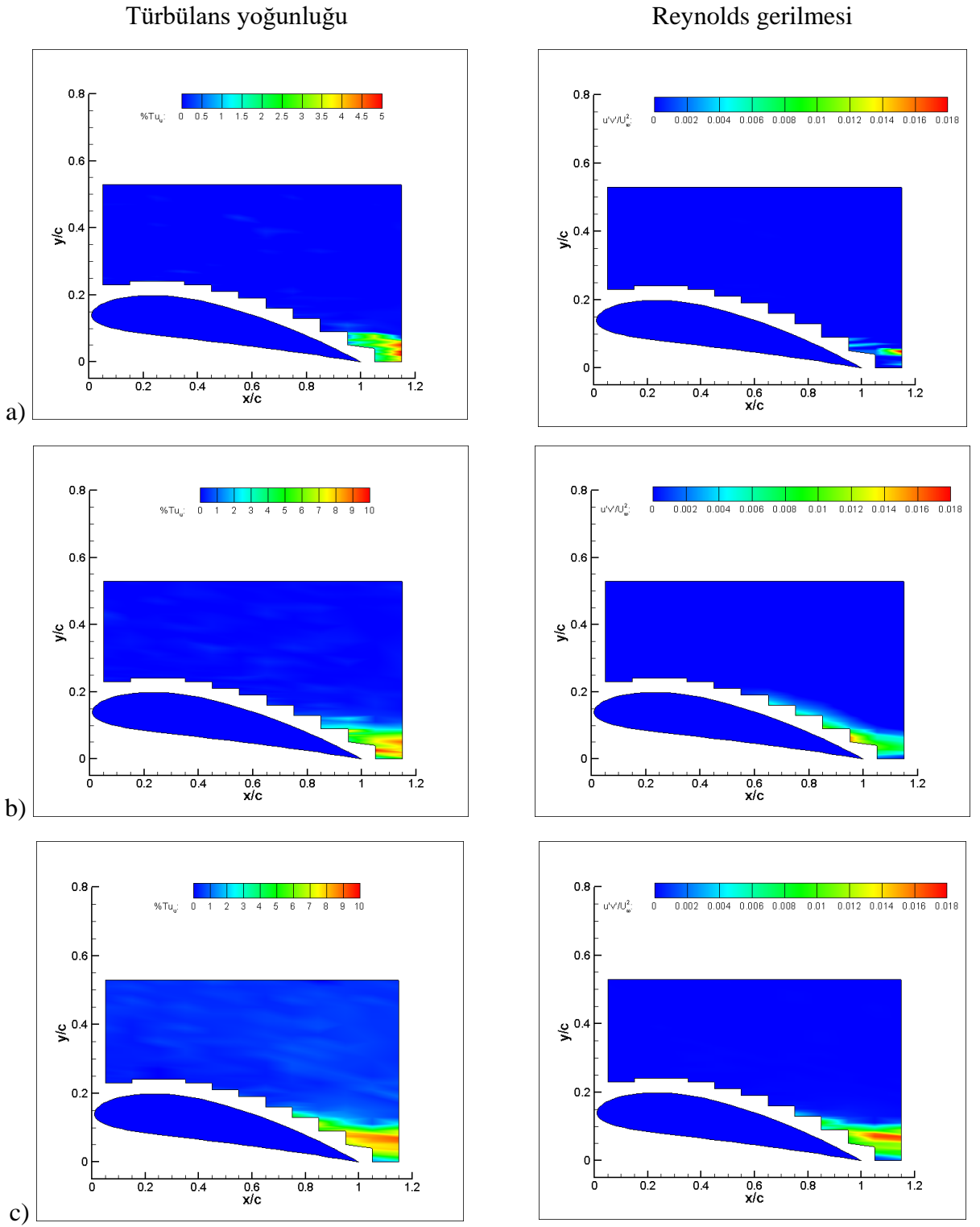

Şekil 3.9b. z/s=0.1'de kanat üzerinde türbülans yoğunluğu ve Reynolds gerilmesi dağılımları α=8° a) Re=25000 c) 50000 b) Re=75000

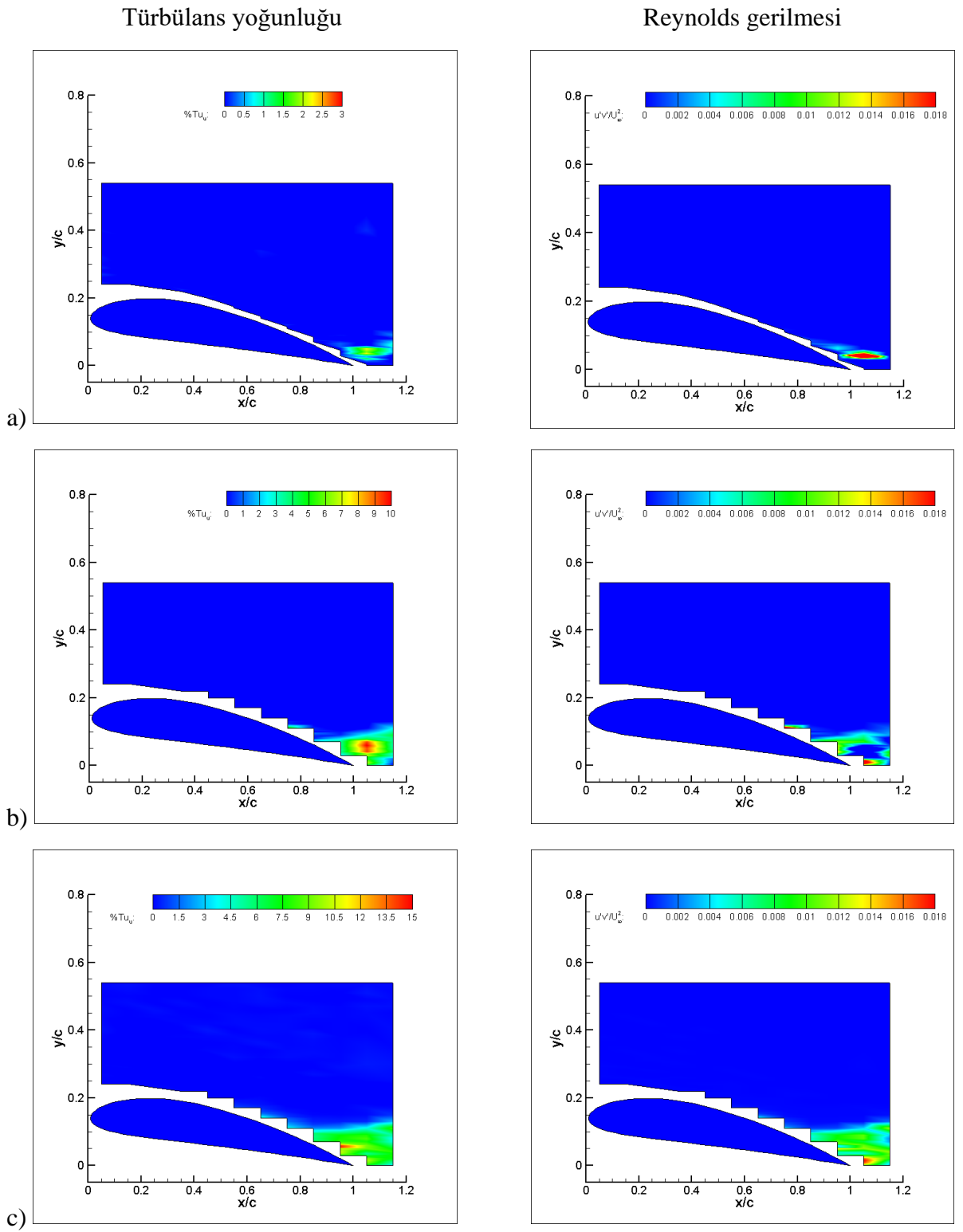

Şekil 3.9c. z/s=0'da kanat üzerinde türbülans yoğunluğu ve Reynolds gerilmesi dağılımları α=8°
a) Re=25000 c) 50000 b) Re=75000

### 3.2.6. Girdap Kopmaları

Düşük Re sayılarında kanadın hücum açısı ile girdap kopma frekanslarının değişimi sabit sıcaklıklı sıcak tel anemometresi ile kanat iz bölgesinde 2 farklı uzaklıkta Şekil 3.10a'da görüldüğü gibi (1C ve 2C) hız ölçümleri yapılmıştır. Bu ölçümlerde elde edilen hızların MATLAB'da Welch fonksiyonu ile spektral analiz sonucu elde edilmiştir. Kanat iz bölgesinde 1C ve 2C uzaklıklarda alınan dikine mesafedeki her bir noktanın spektral analizleri yapılıp bu analiz sonuçları tek tek değerlendirilerek özellikle girdap oluşumunun meydana geldiği bölgelerde ortak oluşan girdap frekanslarından seçme yapılmıştır. Frekans belirlerken tekrarlayan (tünelin frekansı gibi) frekanslar dikkate alınmamıştır. Uç girdapları için de; z/s= - 0.1'de hücum kenarında x/c=0.1'den x/c=1.1'e yani firar kenarı dışına kadar farklı noktalarda ölçümler alınmış ve yine bu ölçümler tek tek değerlendirilmiştir.

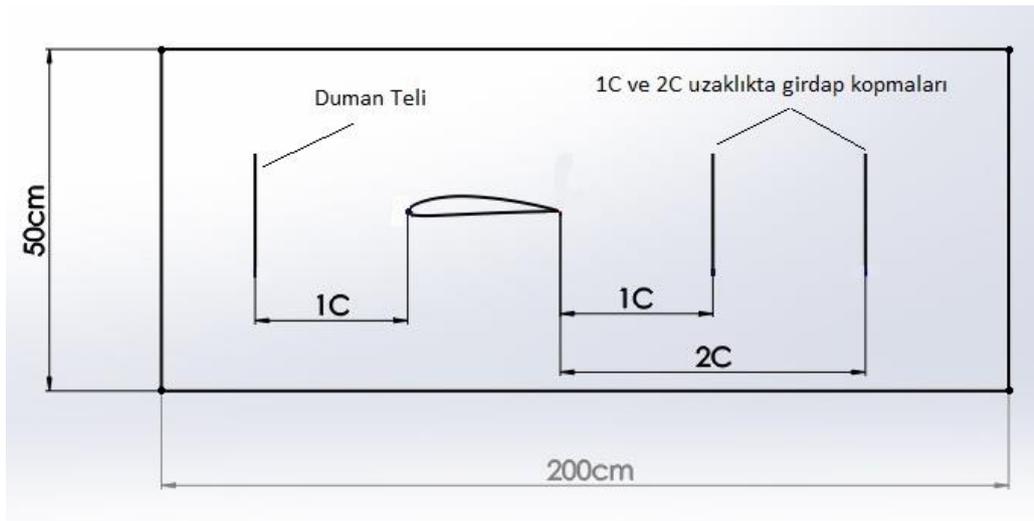

Şekil 3.10a. Duman teli ve girdap kopma deneyi krom-nikel tel ve sıcak tel anemometresi prob konumları

Ölçümler sonucunda değişik Re sayısı ve hücum açılarında elde edilen kopma frekansları sonucu hesaplanan Strouhal (St) sayıları Şekil 3.10b'de verilmiştir. Bu grafikler incelendiğinde; bütün kanatların hücum açısına göre girdap kopma frekansları sonucu elde edilen St sayıları özellikle yüksek hücum açılarında 0.17/sinα şeklinde çizdirilen eğri çevresinde değişmektedir, yani Re sayısı değişimi girdap kopma frekanslarında çok fazla değişikliğe sebep olmamaktadır. Fakat düşük hücum açılarında hücum kenarı ayrılmaları dolayısıyla kopan girdaplar nedeniyle St sayısı değişimleri eğri dışına çıkmaktadır.

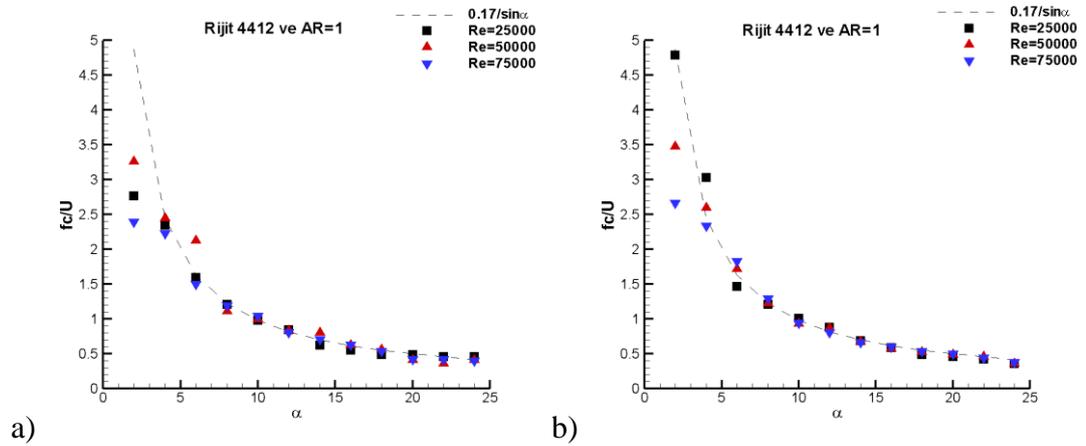

a)                                        b)

Şekil 3.10b. Kanadın iz bölgesinde girdap kopma frekansları a) 1C b) 2C

Uç girdaplarının frekanslarının değişik Re sayısı ve hücum açılarında elde edilen kopma frekansları sonucu hesaplanan St sayıları Şekil 3.11'de verilmiştir. Bu grafikler incelendiğinde; düşük hücum açılarında girdap kopma frekansları sonucu elde edilen St sayıları yüksek iken, hücum açısının artışı ile girdapların büyümesi ile St sayısı azalmakta olduğu görülmektedir.

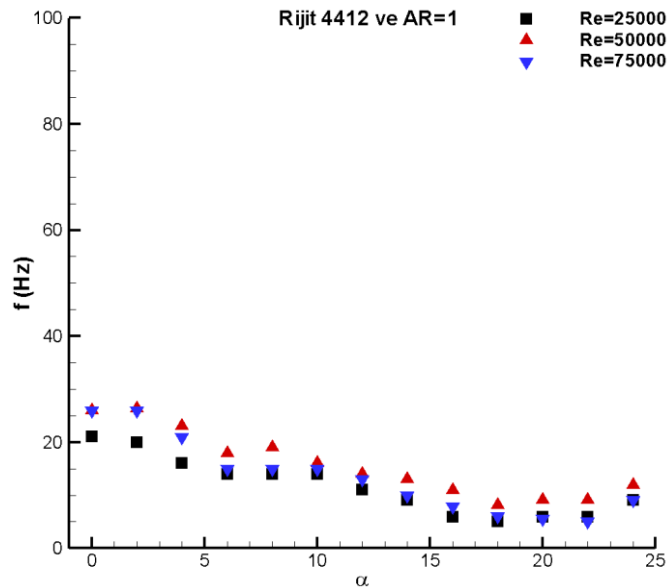

Şekil 3.11. Uç girdap kopma frekanslarının hücum açısına göre değişimleri

### 3.2.7. Duman Teli Deneyi İle Akış Görselleştirme

Duman teli ile akış görselleştirmesinde kanadın hücum kenarının yaklaşık olarak 1 kord boyu önüne gerilmiş krom nikel tel üzerinden akıtılan makine yağının, tele elektrik akımı uygulanması sonucu yanması ile elde edilen duman kullanılmıştır. Krom nikel tel seçiminde öncelikle dikkat edilen husus telin kalınlığı olmuştur. Krom nikel tel çapı akış alanında türbülansa neden olmaması için oldukça küçük olmalıdır. Ancak telin çok ince olması ise telin mukavemeti ve duman kalitesini olumsuz yönde etkileyecektir. Bu yüzden telin Re sayısı yaklaşık olarak 49 olması uygundur [29]. Bu yüzden deney sistemimizde kullanılan telin çapı 0.1 mm olarak seçilmiştir.

Kanat üzerindeki akış görselleştirilmesi öncelikle 0°- 24° arasında 4'er derece adımlarla 25000 ve 50000 Re sayılarında gerçekleştirilmiştir. Akış görselleştirme denemelerimizde tek bir noktadan alınan duman teli görüntüsünün yeterli olmadığı görülmüş ve bu yüzden kanadın uç noktası referans alınarak ($z/s = 0$) kanat merkezine doğru (pozitif yönde) $z/s = +0.1, +0.2$ ve $+0.4$ bölgelerinden geçen duman için görüntüler alınmıştır. Ayrıca uç girdaplarının kanat ucundaki etkisini de iyi bir şekilde anlamak için referans noktası yani $z/s = 0$ ve kanat dışına doğru (negatif yönde) $z/s = -0.1$ ve $-0.2$ bölgeleri içinde duman teli deneyleri yapılmıştır. Duman görüntülerinin sembolik olarak kanadın hangi bölgesinden geçtiği Şekil 3.12'de gösterilmiştir.

Yapılan bu çalışmalardan sonra akışın kanat üstünden ve kanat altından geçişlerini görselleştirebilmek amacı ile Şekil 3.13'deki düzenek kurulmuştur. Bu çalışma ile akışın tamamı görselleştirilerek akışın bütünü için bilgi edinmeye çalışılmıştır. Kanadın yüzeylerinden geçen akışın görselleştirilmesinde krom nikel tel kanat açıklığı eksenine paralel olarak hücum kenarı hizasına denk gelecek şekilde konumlandırılmıştır. Kanadın alt kenarından geçen akışı görselleştirmek için krom nikel tel hücum kenarı hizasından 1mm kaydırılarak duman flementlerinin kanat altından geçmesi sağlanmış böylelikle akış görselleştirilmesi gerçekleştirilmiştir. Aynı şekilde kanat üstündeki akışı görselleştirmek için bu sefer tel kanat üst yüzey yönünde 1mm duman flamentlerinin kanat üzerinden geçmesi sağlanmıştır.

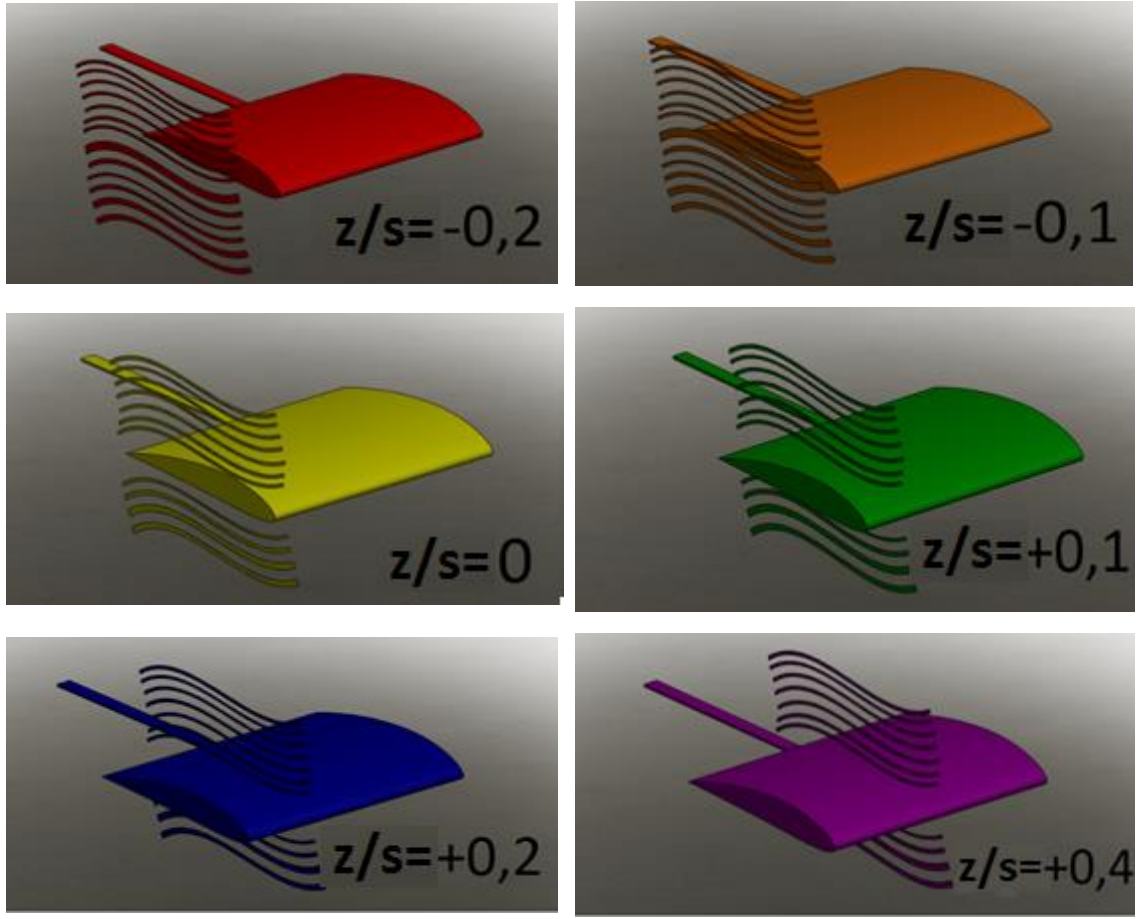

Şekil 3.12. Duman telinden gelen akışın farklı konumlardan geçişi
z/s = -0.2, -0.1, 0, +0.1, +0.2, +0.4

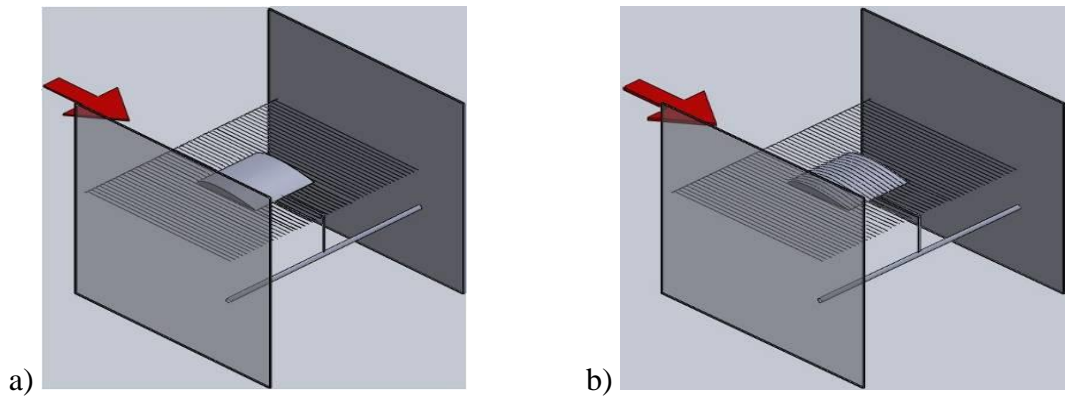

Şekil 3.13. Duman telinden gelen akışın kanadın a) alt ve b) üst yüzeyinden geçişi

Re=25000'de düşük hücum açılarında firar kenarına yakın ayrılma kabarcıkları oluşmakta ve bu kabarcıklar hücum açısı arttıkça hem enine ve boyuna küçülmekte hem de hücum kenarına doğru hareket etmektedir.

Re=25000'de orta hücum açılarında (12º-16º) ayrılma kabarcığı kısa ayrılma kabarcığından uzun ayrılma kabarcığına dönüşmekte ve özellikle 16º'de uzun ayrılma kabarcığı patlayarak kanat üzerinde dağılmaktadır. Dağılan bu kabarcık kanat iz bölgesine kadar devam etmekte ve Re sayısı düşük olduğu için atalet kuvvetleri fazla olmadığından iz bölgesinden kanada doğru akışı kararsız yapan ters akış oluşturmaktadır (Şekil 3.14-3.19). Bu geri doğru akış özellikle Re=50000 olduğunda atalet kuvvetleri biraz daha fazla olduğu için görülmemekte ve akıştaki kararsızlık daha az olmaktadır (Şekil 3.20-3.25). Re=75000'de bu tür durumun daha az olduğu ve kuvvet katsayısı grafiklerinde de görüldüğü gibi aerodinamik performansların iyileştiği sonucu rahatlıkla çıkarılabilmektedir.

Kanat için ayrılma kabarcıkları oluşan uç girdaplarından etkilenmekte, girdaplar akışı tutundurmaya çalışan bir ilave kuvvet gibi davranmakta ve ortadaki ayrılmış akış uçlarda tutunmuş akışa dönmektedir. Kanat üst yüzeyinin çoğuna her iki uçtan gelen girdaplar etki etmekte ve bu girdaplar kanat iz bölgesinde dönmeli akış oluşturarak (özellikle 25000 Re sayısında) kararsızlık artmakta ve ters akışlar oluşmaktadır. Kanat iz bölgesindeki dönümlü akış bölgesi basınç düşüşü ile vakum etkisi yaparak kanat üzerindeki akışı iyileştirme etkisi yapar ve tutunma kaybı aniden oluşmadan yumuşak tutunma kaybı ile akış daha geç tutunma kaybına uğrar.

Uç girdapları düşük hücum açılarında kanat uçlarında oluşup iz bölgesine doğru devam ederken hücum açısının artışı ile yatay olarak hareket etmekte ve kanadın içlerine doğru yönelmektedir. Bu durumun her iki uç girdabının birbirine etki ettiği (kanat üzerinde dönmeli ve daha fazla enerjili akış oluşturdukları için) ve bu etkiler ile ortalara kadar (yatay) hareket edebildikleri düşünülmektedir.

Ayrıca 15º-25º orta hücum açılarında kuvvet katsayısı grafiklerinde görülen düşmelerin o hücum açılarında hücum kenarı ayrılmalarının baskın olmasından kaynaklandığı duman teli deneylerinden görülmektedir (Şekil 3.14-3.29).

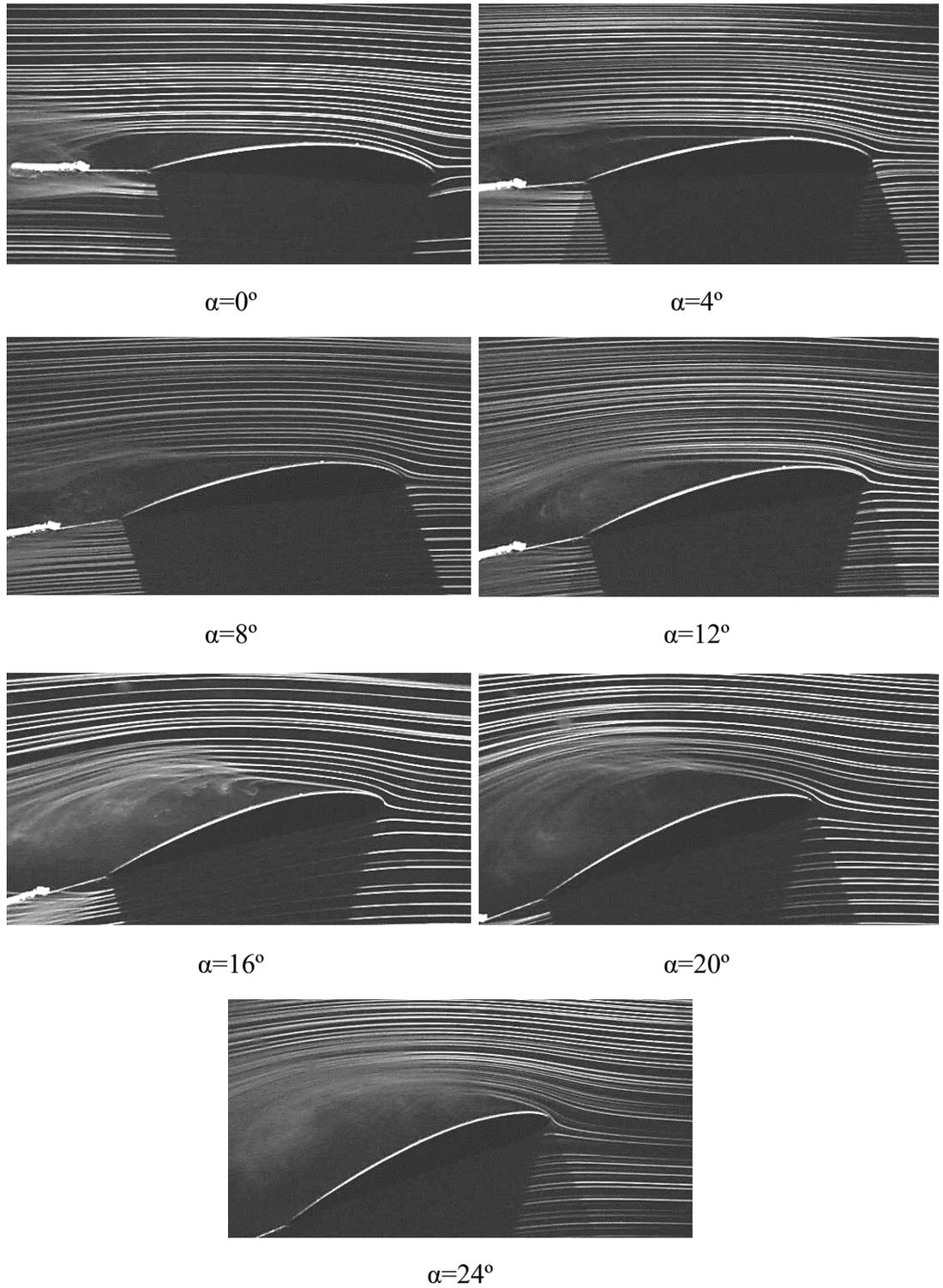

α=0°  α=4°

α=8°  α=12°

α=16°  α=20°

α=24°

Şekil 3.14. Duman teli ile akış görselleştirme (Re=25000 z/s=0.4)

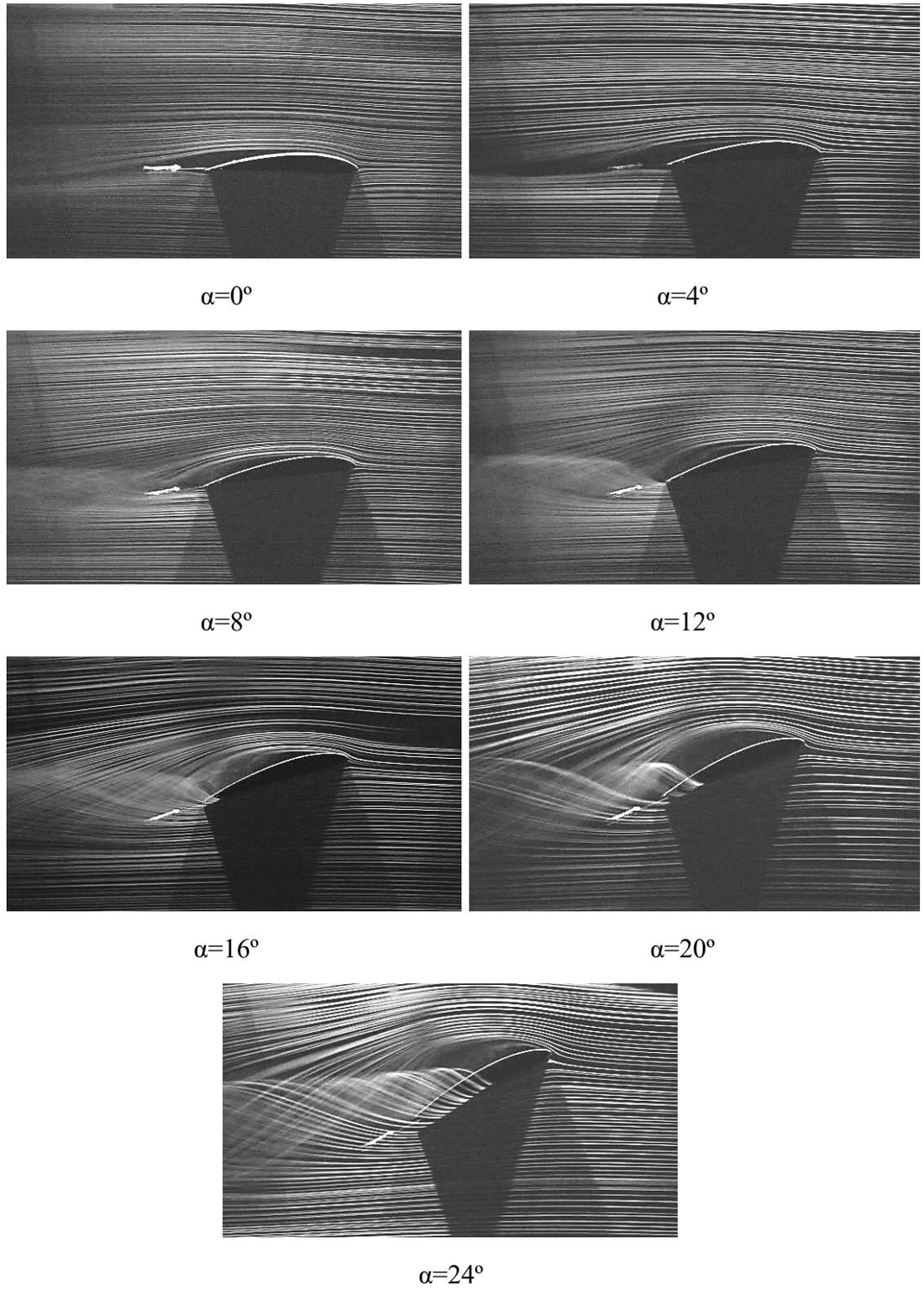

α=0º        α=4º

α=8º        α=12º

α=16º        α=20º

α=24º

Şekil 3.15. Duman teli ile akış görselleştirme (Re=25000 z/s=0.2)

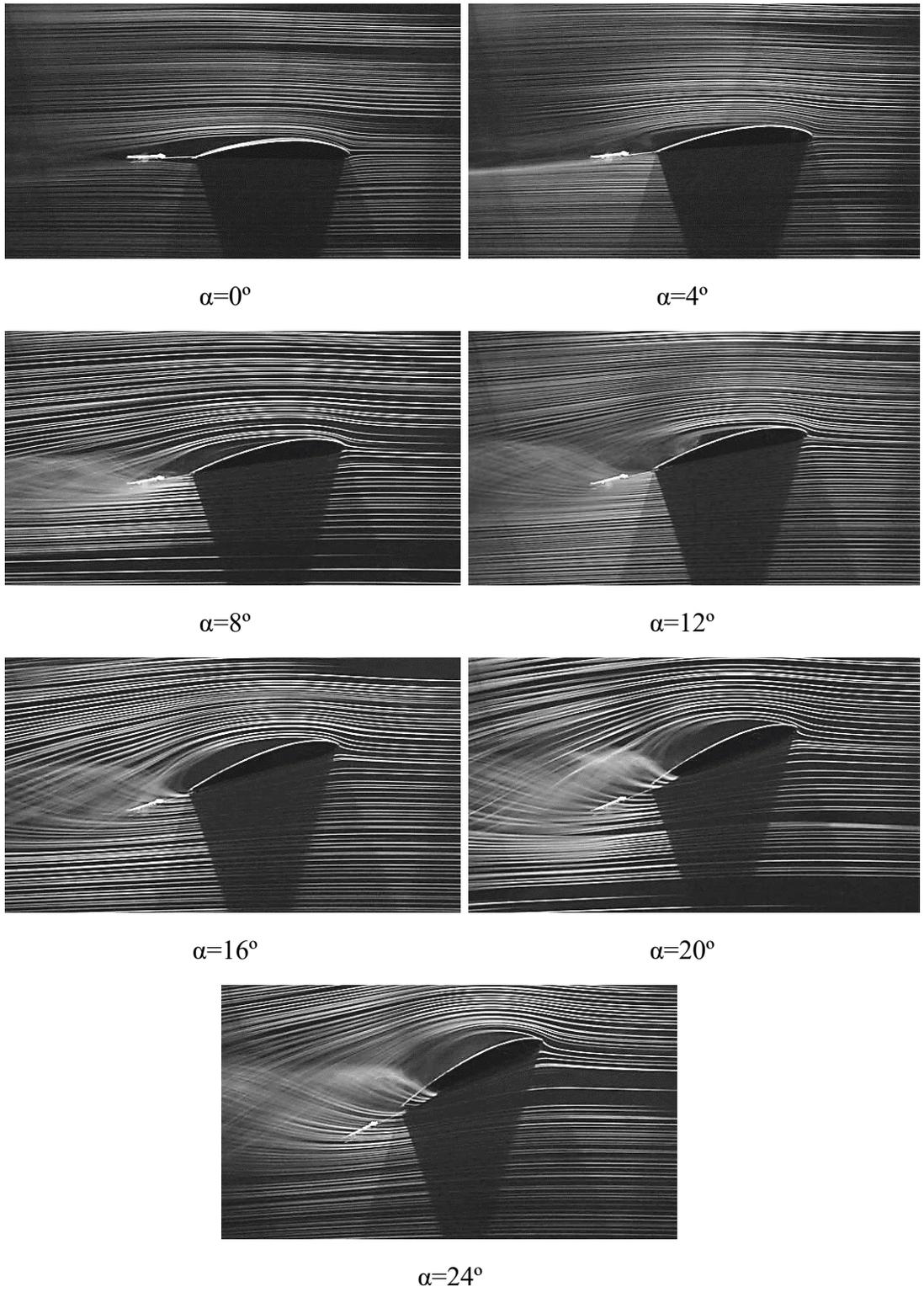

$\alpha=0^\circ$      $\alpha=4^\circ$

$\alpha=8^\circ$      $\alpha=12^\circ$

$\alpha=16^\circ$      $\alpha=20^\circ$

$\alpha=24^\circ$

Şekil 3.16. Duman teli ile akış görselleştirme (Re=25000 z/s=0.1)

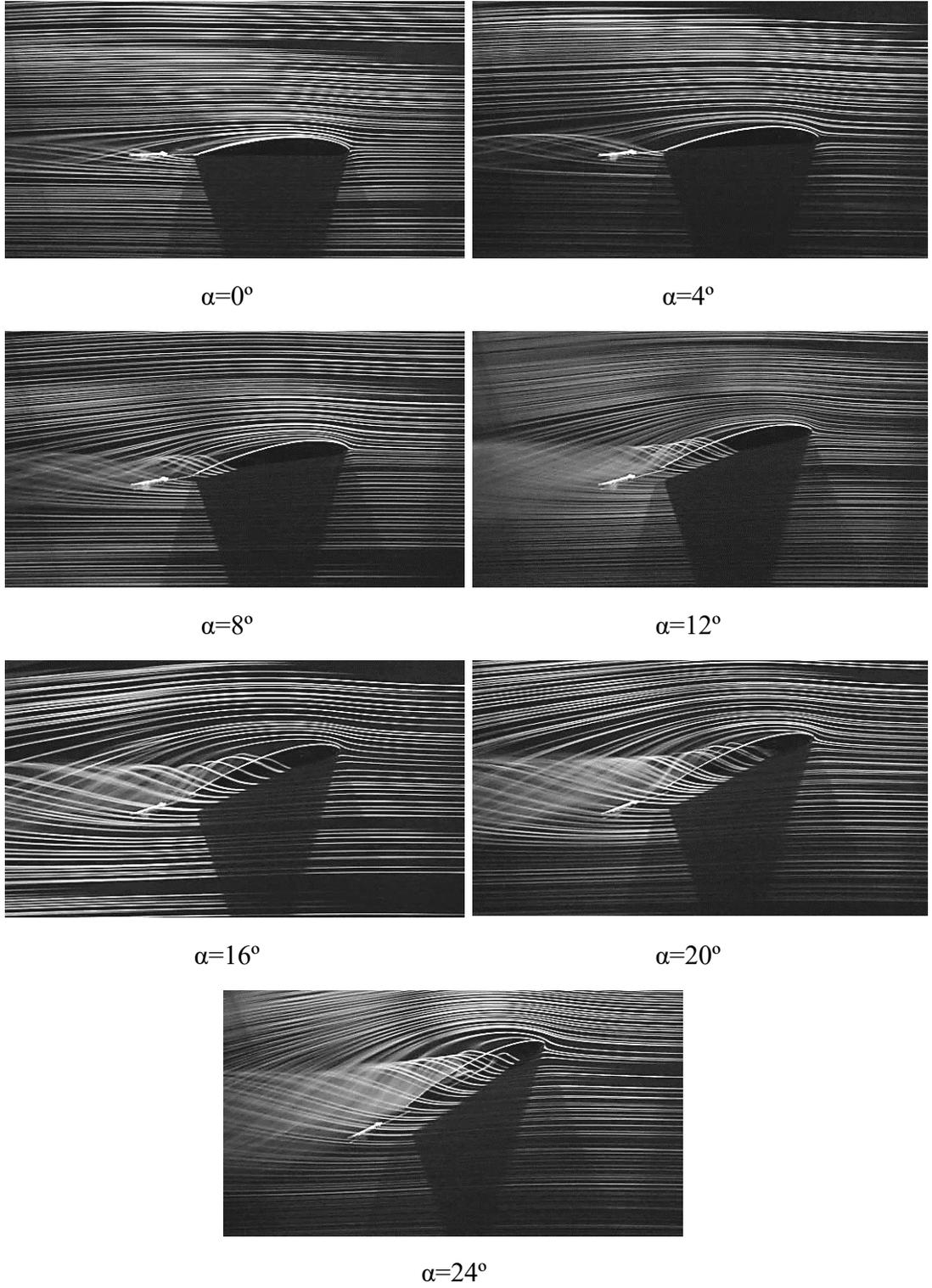

α=0º    α=4º

α=8º    α=12º

α=16º    α=20º

α=24º

Şekil 3.17. Duman teli ile akış görselleştirme (Re=25000 z/s=0)

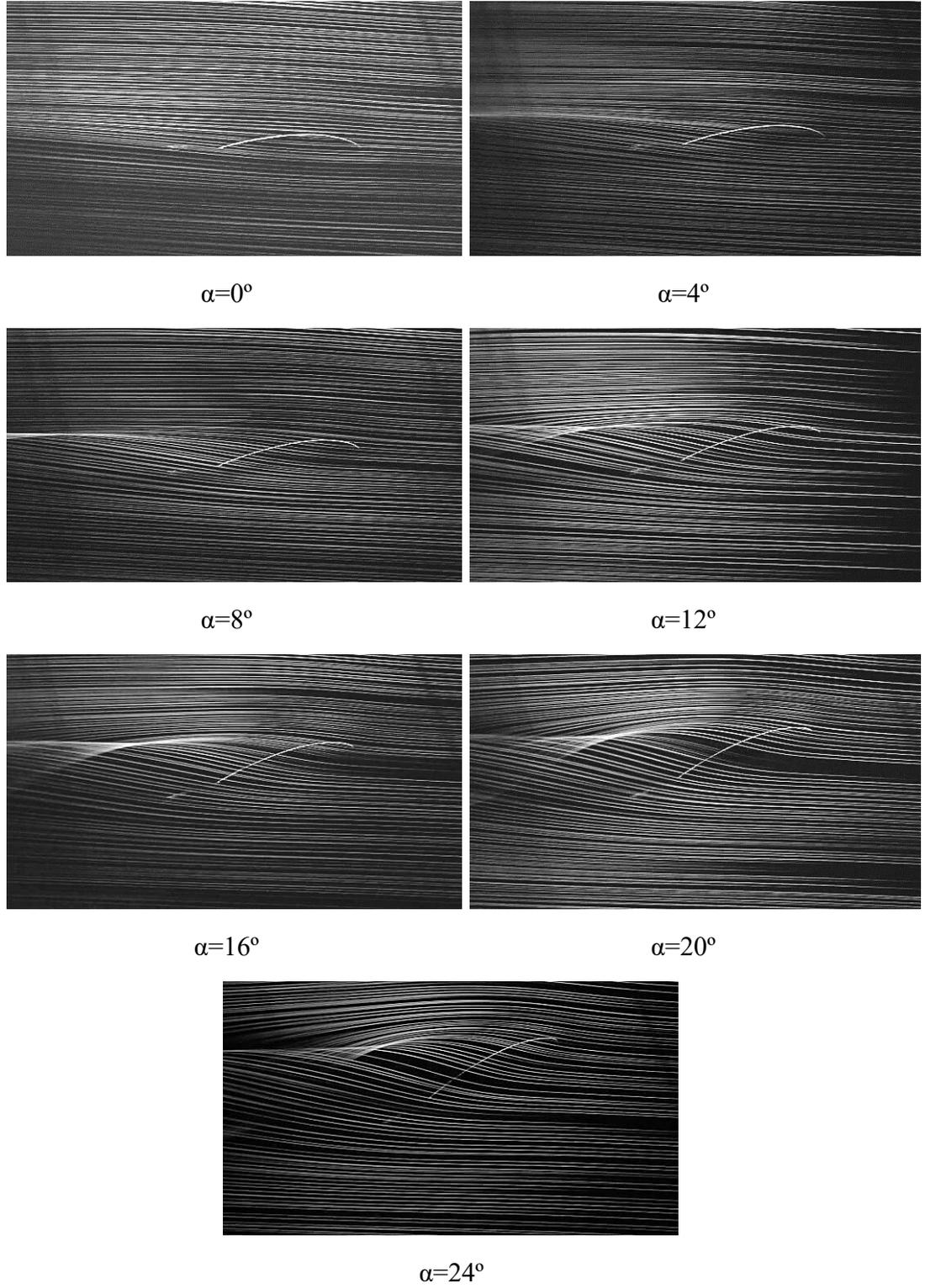

α=0º                    α=4º

α=8º                    α=12º

α=16º                   α=20º

α=24º

Şekil 3.18. Duman teli ile akış görselleştirme (Re=25000 z/s=-0.1)

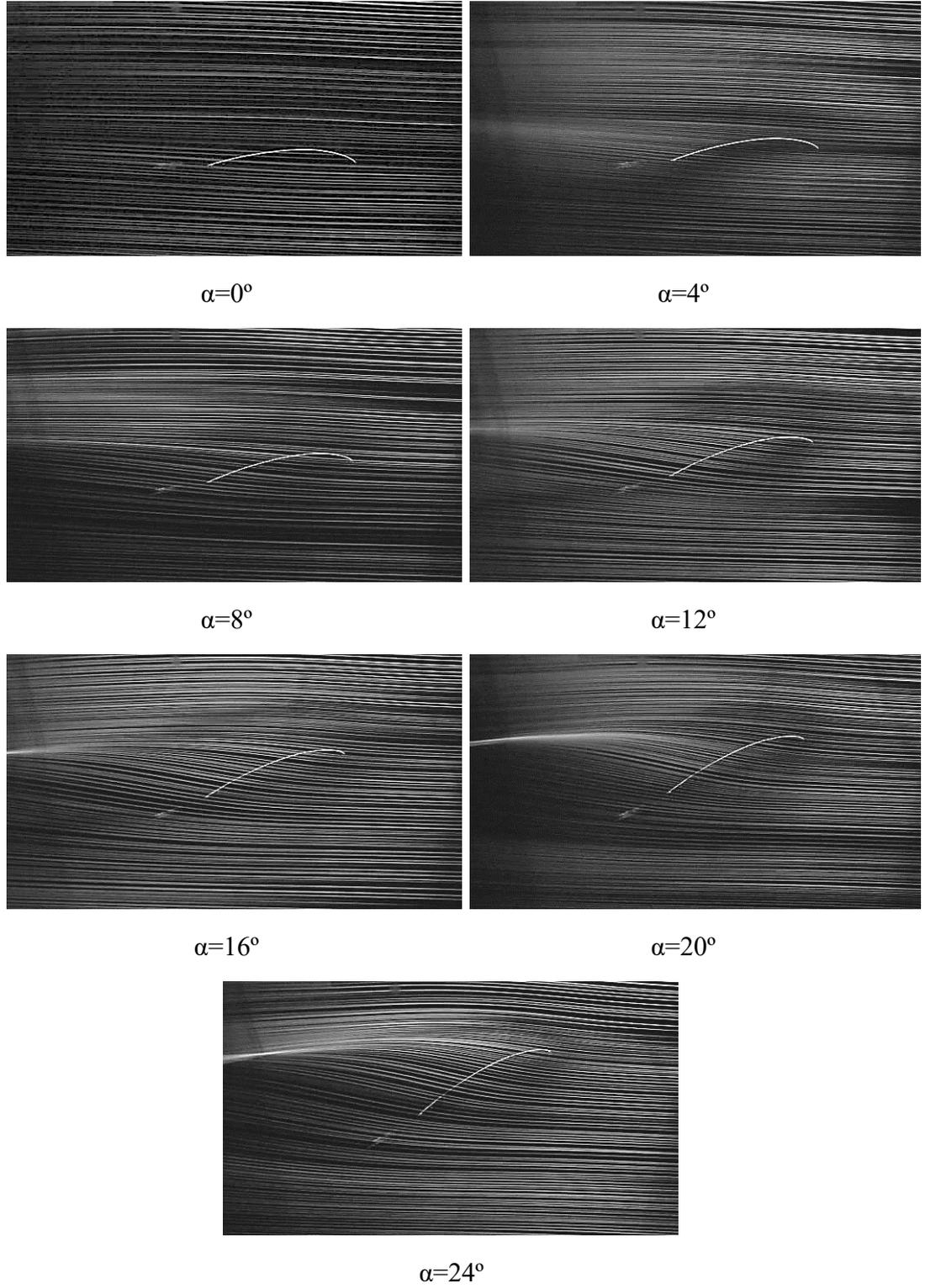

α=0º                        α=4º

α=8º                      α=12º

α=16º                     α=20º

α=24º

Şekil 3.19. Duman teli ile akış görselleştirme (Re=25000 z/s=-0.2)

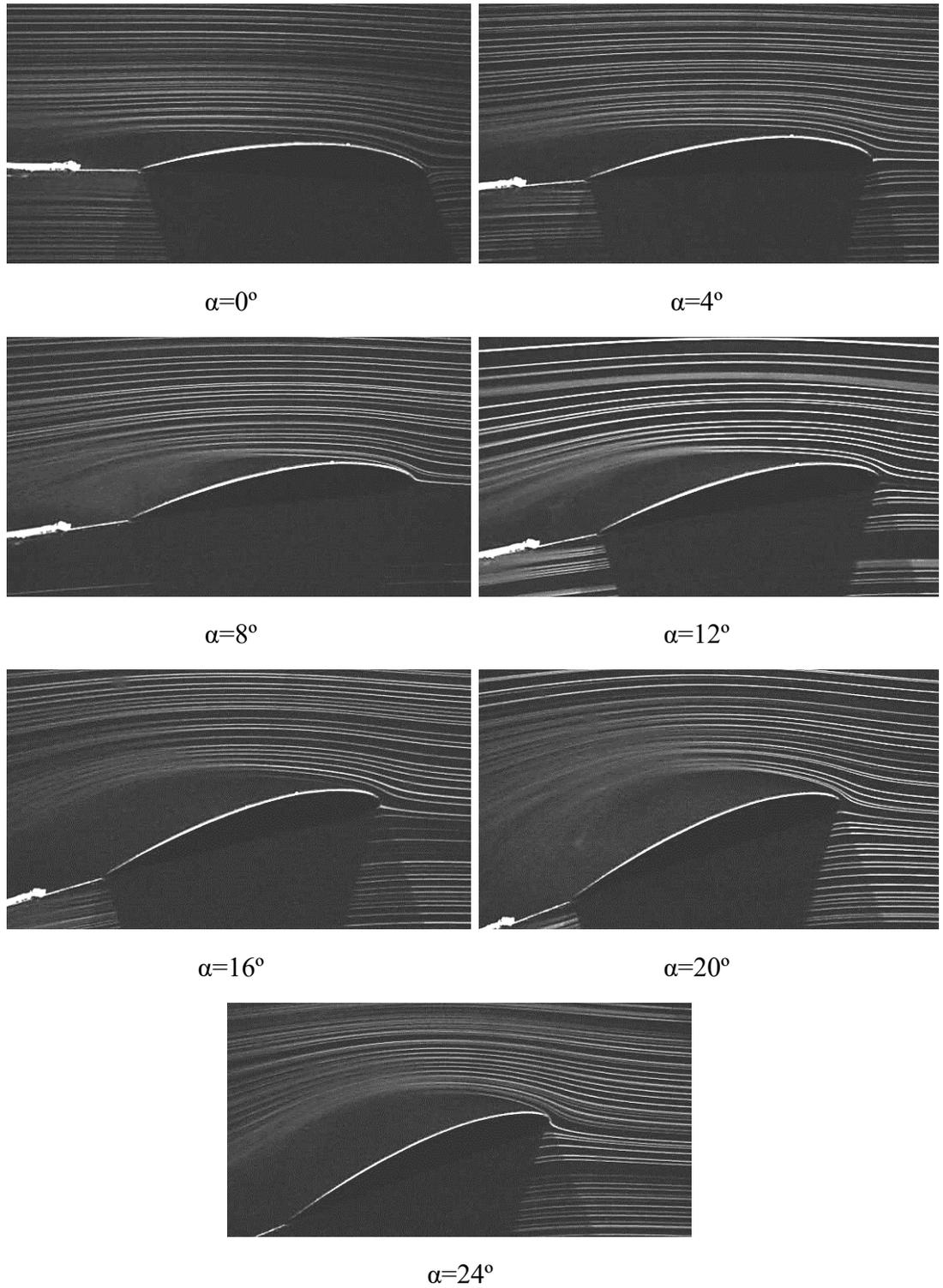

α=0º    α=4º

α=8º    α=12º

α=16º    α=20º

α=24º

Şekil 3.20. Duman teli ile akış görselleştirme (Re=50000 ve z/s=0.4)

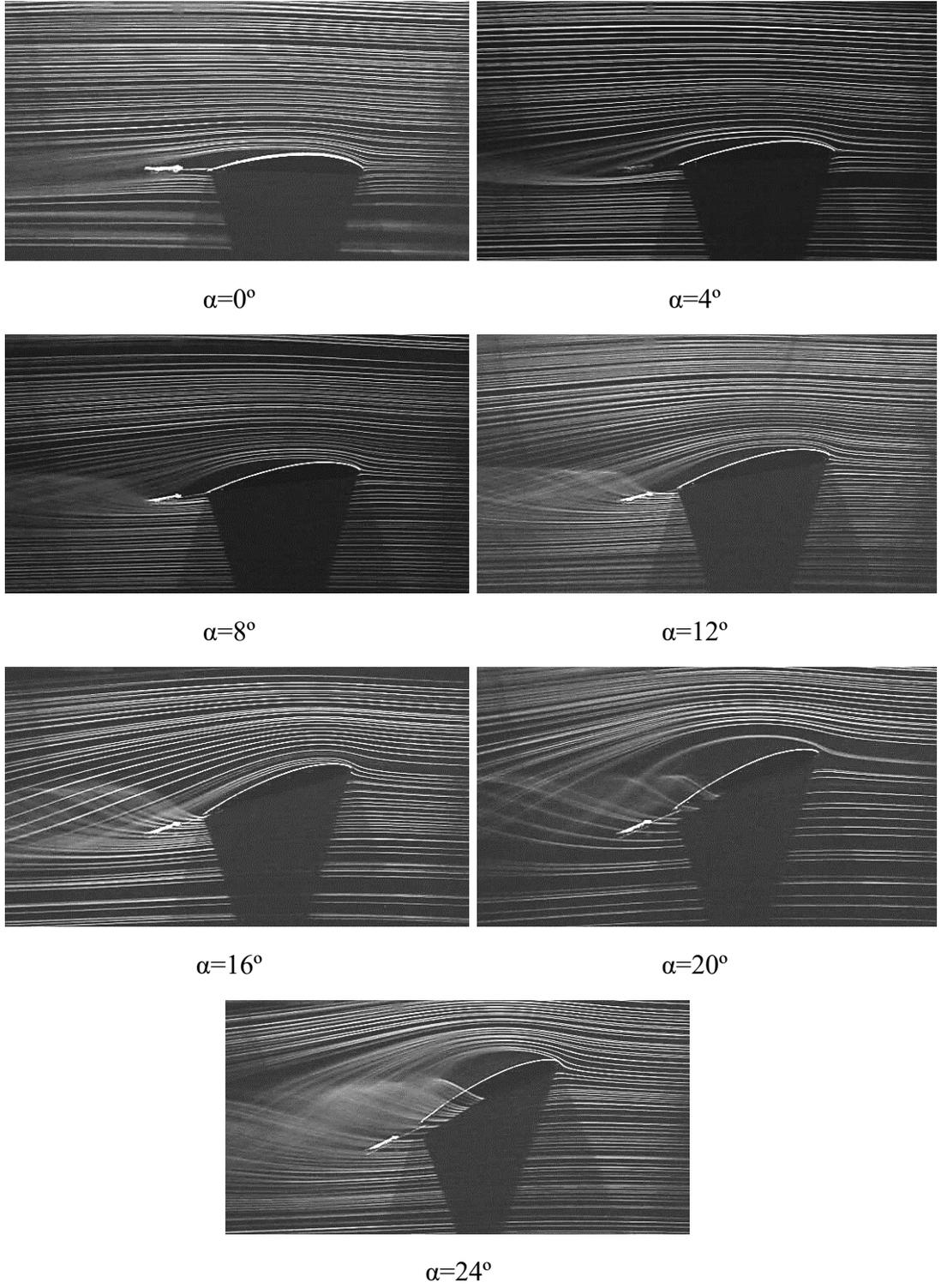

α=0°                              α=4°

α=8°                              α=12°

α=16°                             α=20°

α=24°

Şekil 3.21. Duman teli ile akış görselleştirme (Re=50000 z/s=0.2)

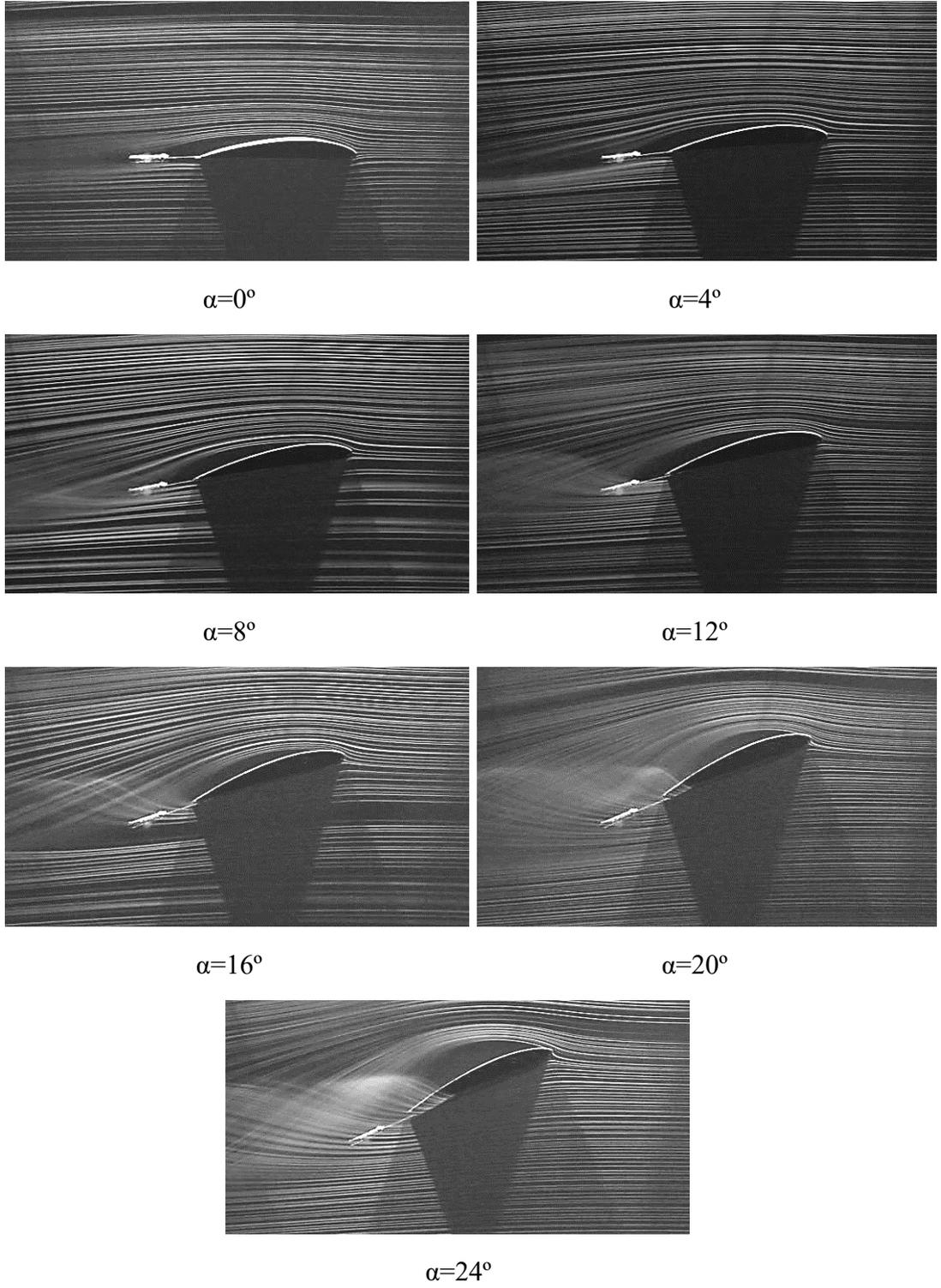

α=0°  α=4°

α=8°  α=12°

α=16°  α=20°

α=24°

Şekil 3.22. Duman teli ile akış görselleştirme (Re=50000 z/s=0.1)

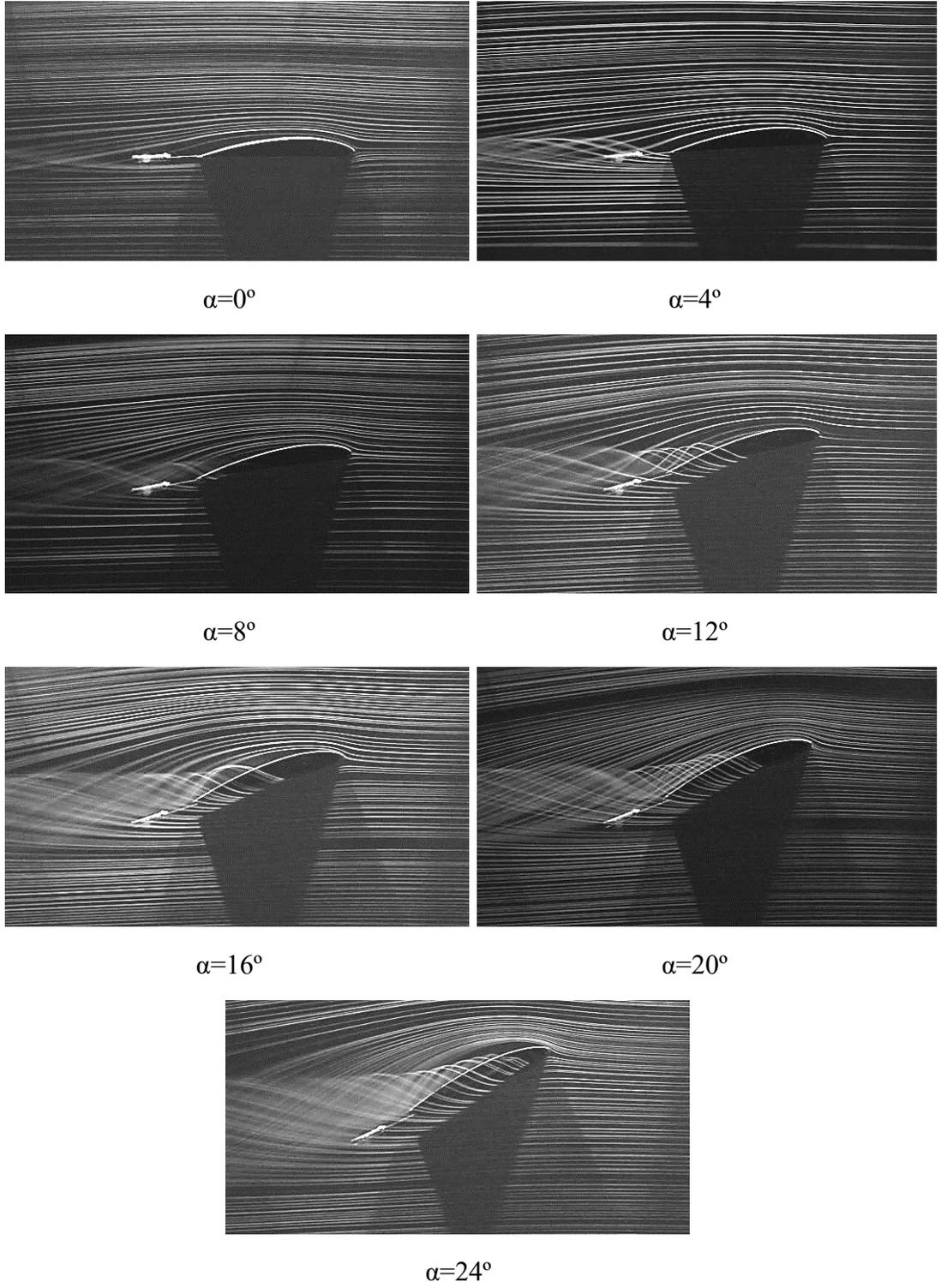

α=0º  α=4º

α=8º  α=12º

α=16º  α=20º

α=24º

Şekil 3.23. Duman teli ile akış görselleştirme (Re=50000 z/s=0)

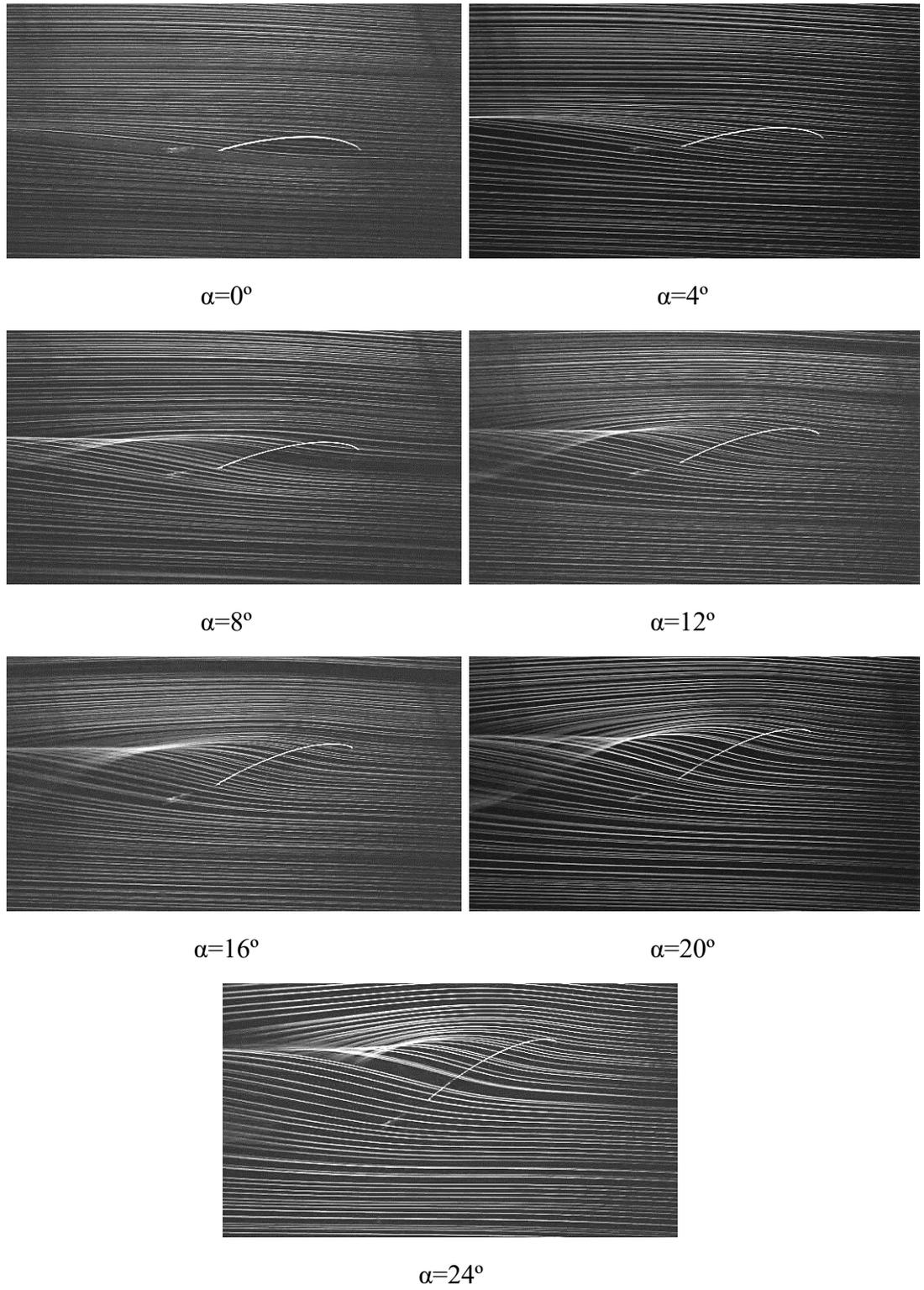

α=0º            α=4º

α=8º            α=12º

α=16º            α=20º

α=24º

Şekil 3.24. Duman teli ile akış görselleştirme (Re=50000 z/s=-0.1)

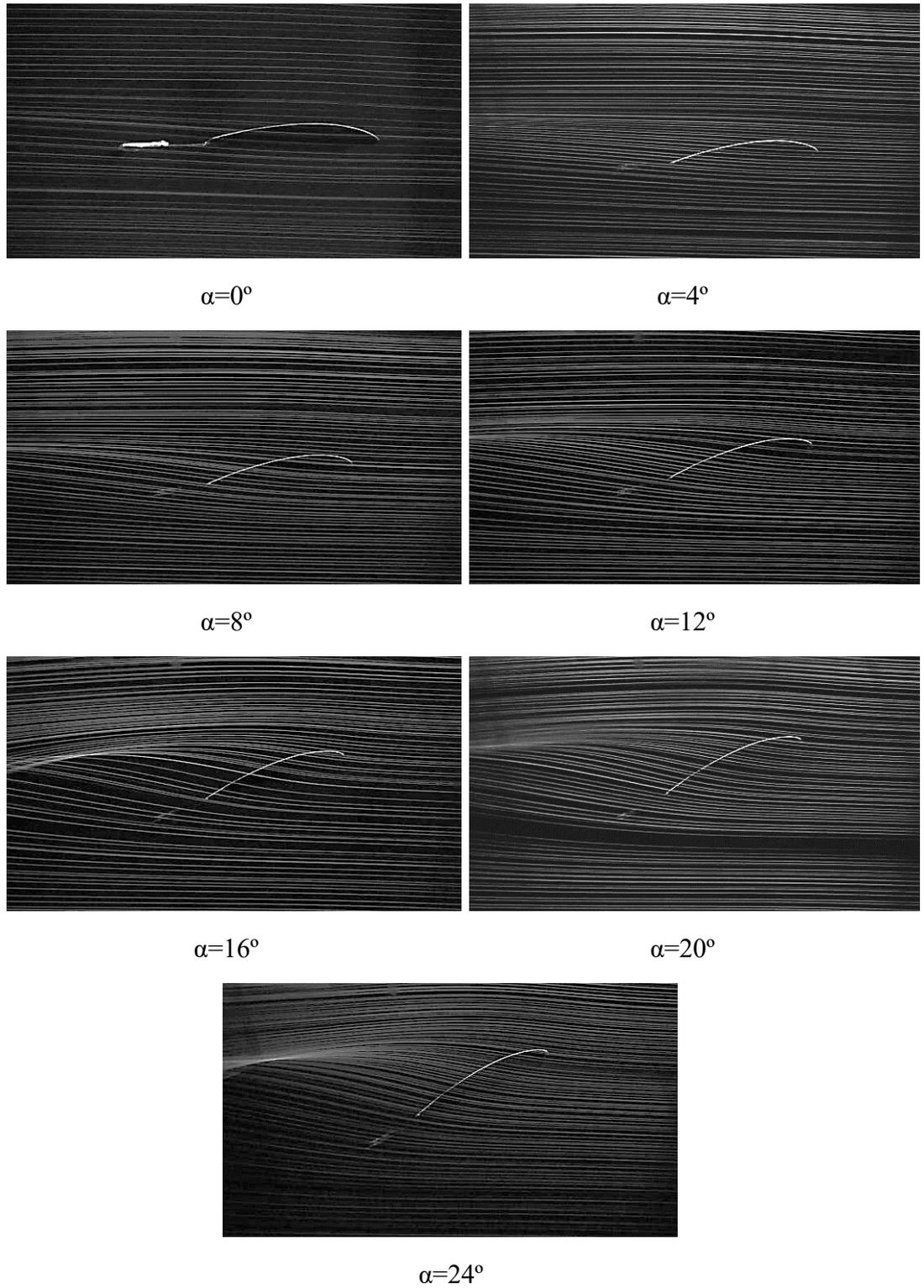

$\alpha=0^{\circ}$        $\alpha=4^{\circ}$

$\alpha=8^{\circ}$        $\alpha=12^{\circ}$

$\alpha=16^{\circ}$        $\alpha=20^{\circ}$

$\alpha=24^{\circ}$

Şekil 3.25. Duman teli ile akış görselleştirme (Re=50000 z/s=-0.2)

Ancak yapılan deney sonuçlarının incelenmesi sonucunda akış ayrılmasının belirlenmesi kesin olarak mümkün olmamıştır. Bu yüzden duman ile akış görselleştirme deneyleri Re=25000 ve 30°, 35° ve 40° derece hücum açıları ve z/s = +0.4 ve z/s = +0.1 için tekrar yapılmıştır. Şekil 3.26 ve 3.27'den de görüldüğü gibi kanat ortalarında akış tamamen ayrılırken kanat ucunda girdap çapları açı ile birlikte büyümektedir.

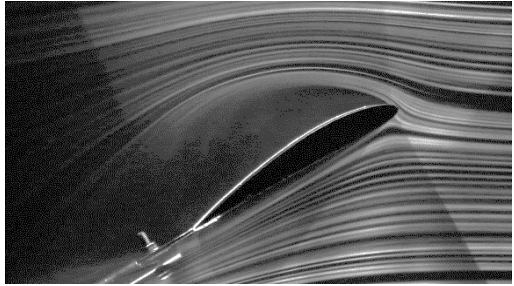
α=30°

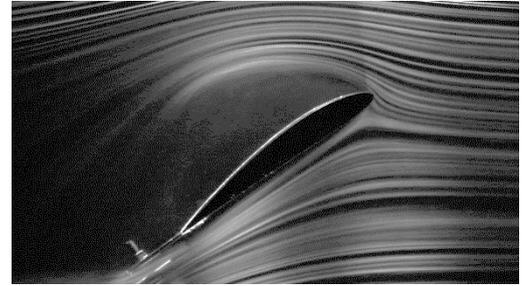
α=35°

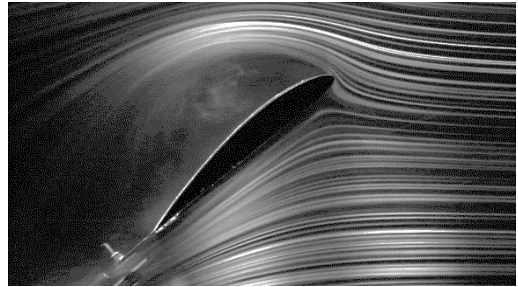
α=40°

Şekil 3.26. Duman teli ile akış görselleştirme (Re=25000 z/s=0.4)

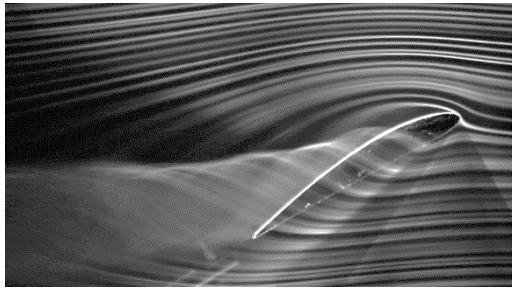
α=30°

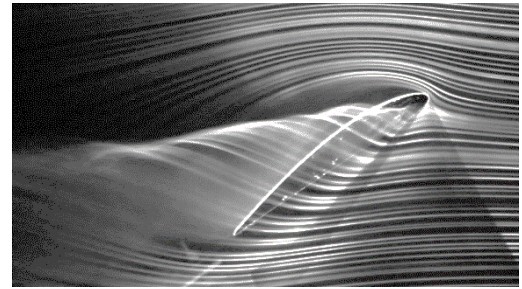
α=35°

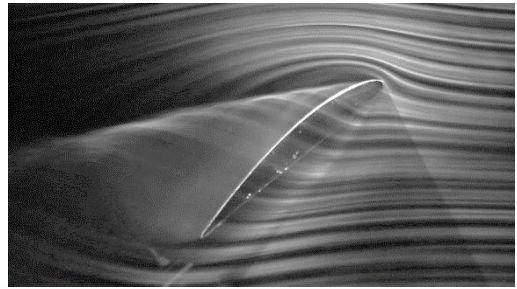
α=40°

Şekil 3.27. Duman teli ile akış görselleştirme (Re=25000 z/s=0.1)

Duman teli ile akış görselleştirme deneylerinden de anlaşılacağı üzere düşük hücum açılarında uç girdapları kanat firar kenarlarında oluşmakta ve hücum açısının artması ile uç girdap kopmaları kanadın hücum kenarına doğru kaymaktadır. Ayrıca hücum açısının artması ile birlikte uç girdapların çapı ve kanat yüzeyine etkisi de arttığı Şekil 3.28 ve 3.29'da görülmektedir. Re=25000'de ve AoA=0° olduğunda uç girdapları kanat üst yüzeyinde firar kenarına yakın bölgelerde görülmektedir. AoA=24° olduğunda uç girdapları kanadın üst bölgesinin yaklaşık olarak yarısını kaplamış durumdadır. Duman teli deneyi ve kuvvet ölçümlerinden de görüldüğü gibi bu girdapların kanat üstündeki akışın ayrılmasını geciktirdiği görülmüştür.

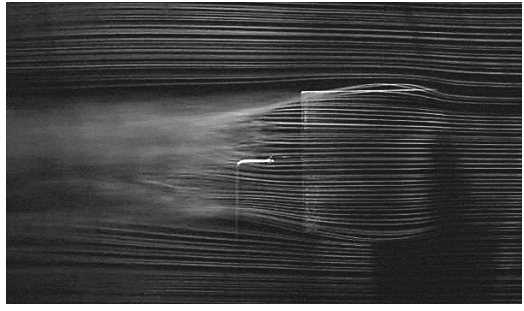 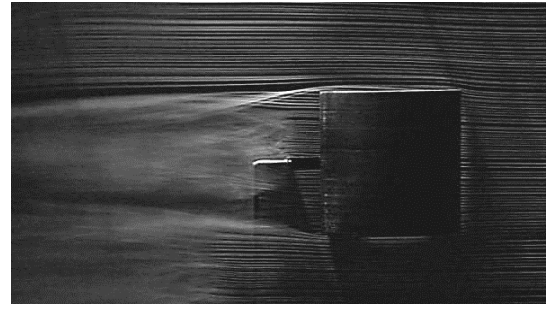

α=0º üst      α=0º alt

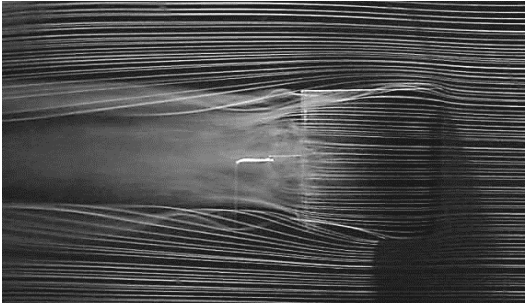 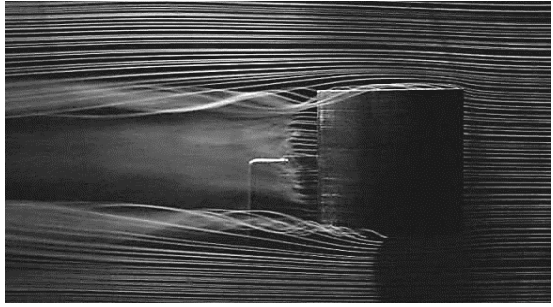

α=4º üst      α=4º alt

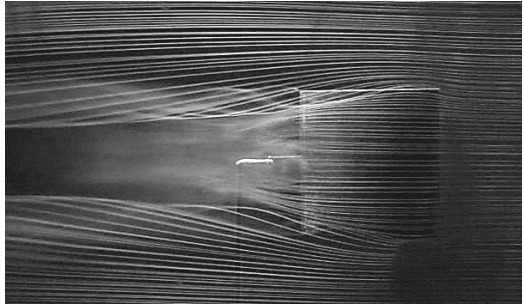 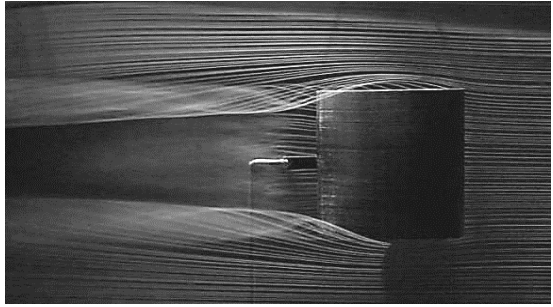

α=8º üst      α=8º alt

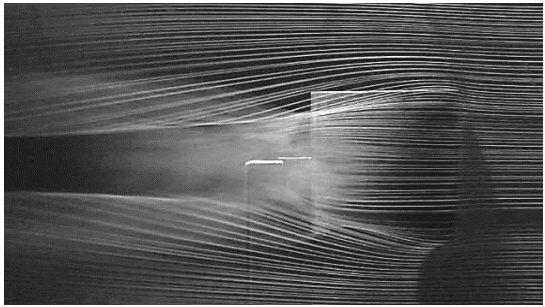 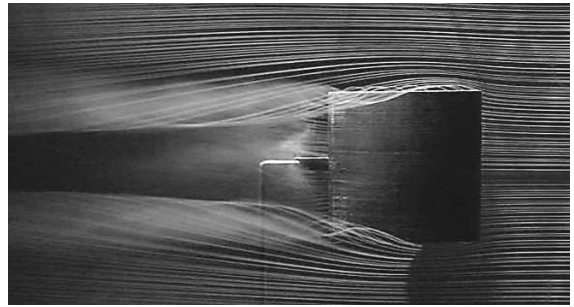

α=12º üst      α=12º alt

Şekil 3.28. Duman teli ile akış görselleştirme Re=25000

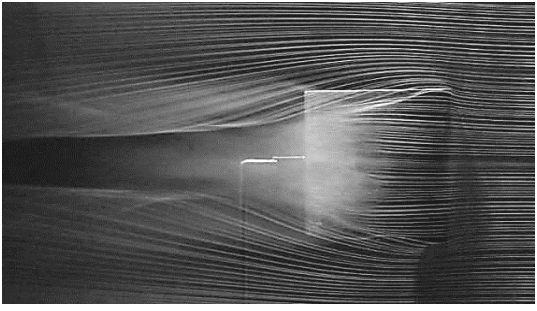 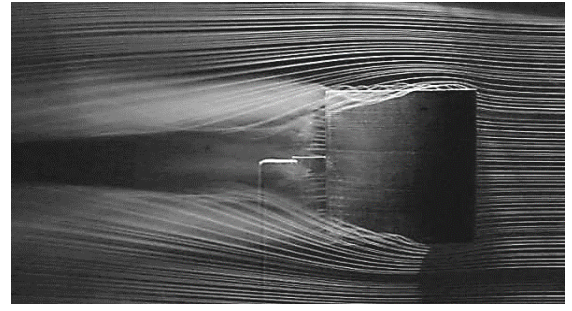

<div align="center">α=16º üst       α=16º alt</div>

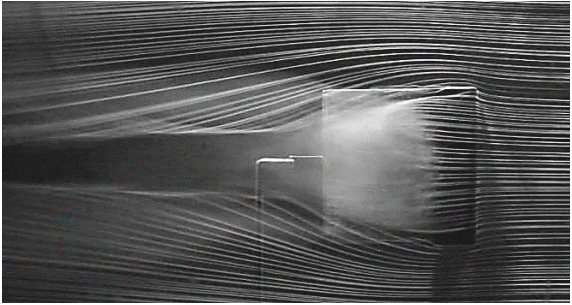 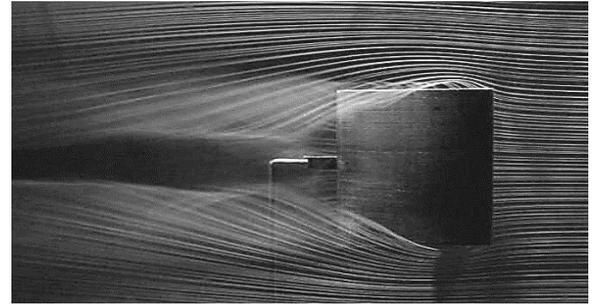

<div align="center">α=20º üst       α=20º alt</div>

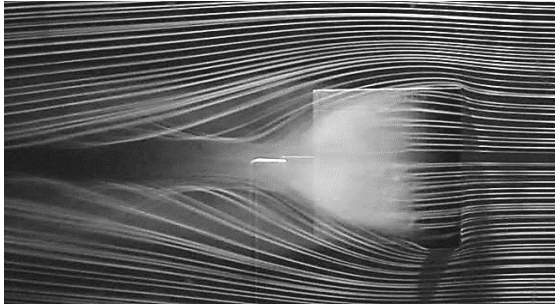 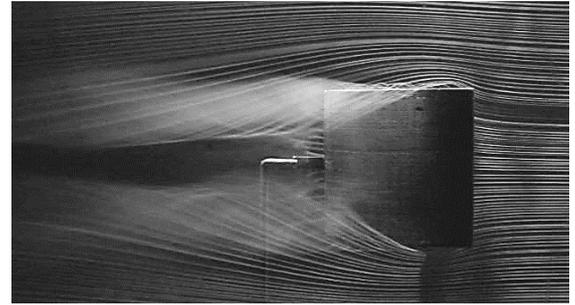

<div align="center">α=24º üst       α=24º alt</div>

Şekil 3.28. Duman teli ile akış görselleştirme Re=25000 (Devamı)

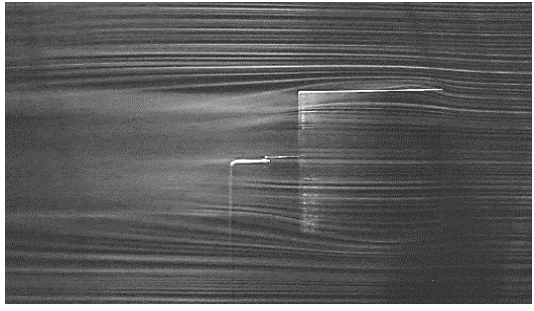 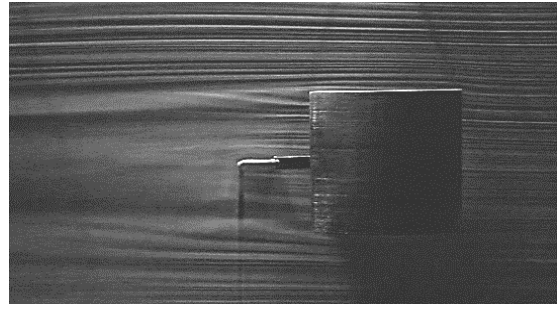

α=0º üst                α=0º alt

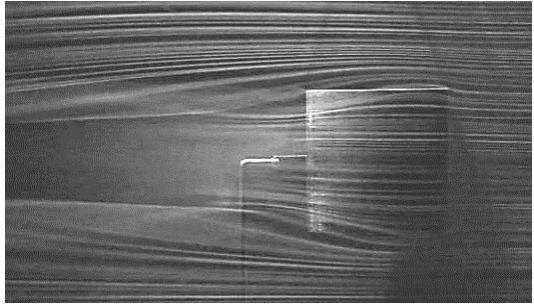 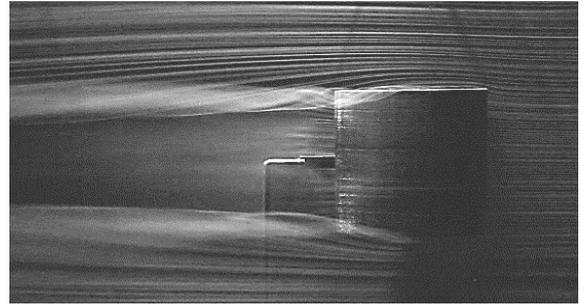

α=4º üst                α=4º alt

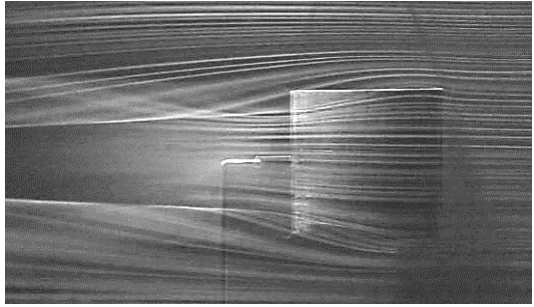 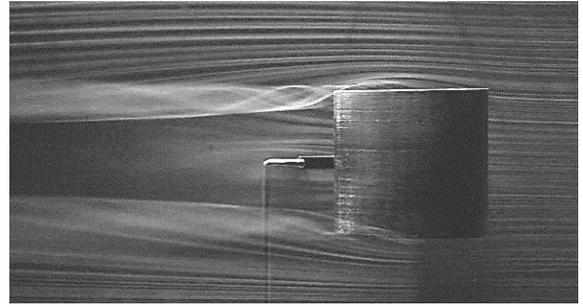

α=8º üst                α=8º alt

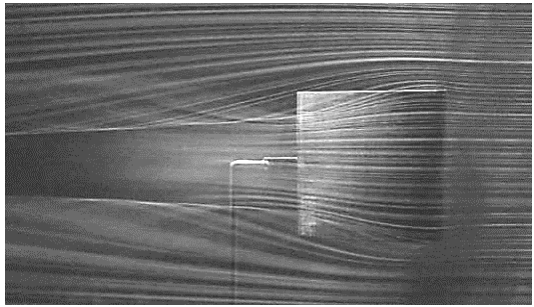 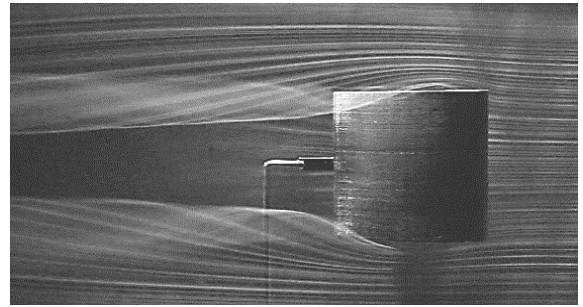

α=12º üst              α=12º alt

Şekil 3.29. Duman teli ile akış görselleştirme Re=50000

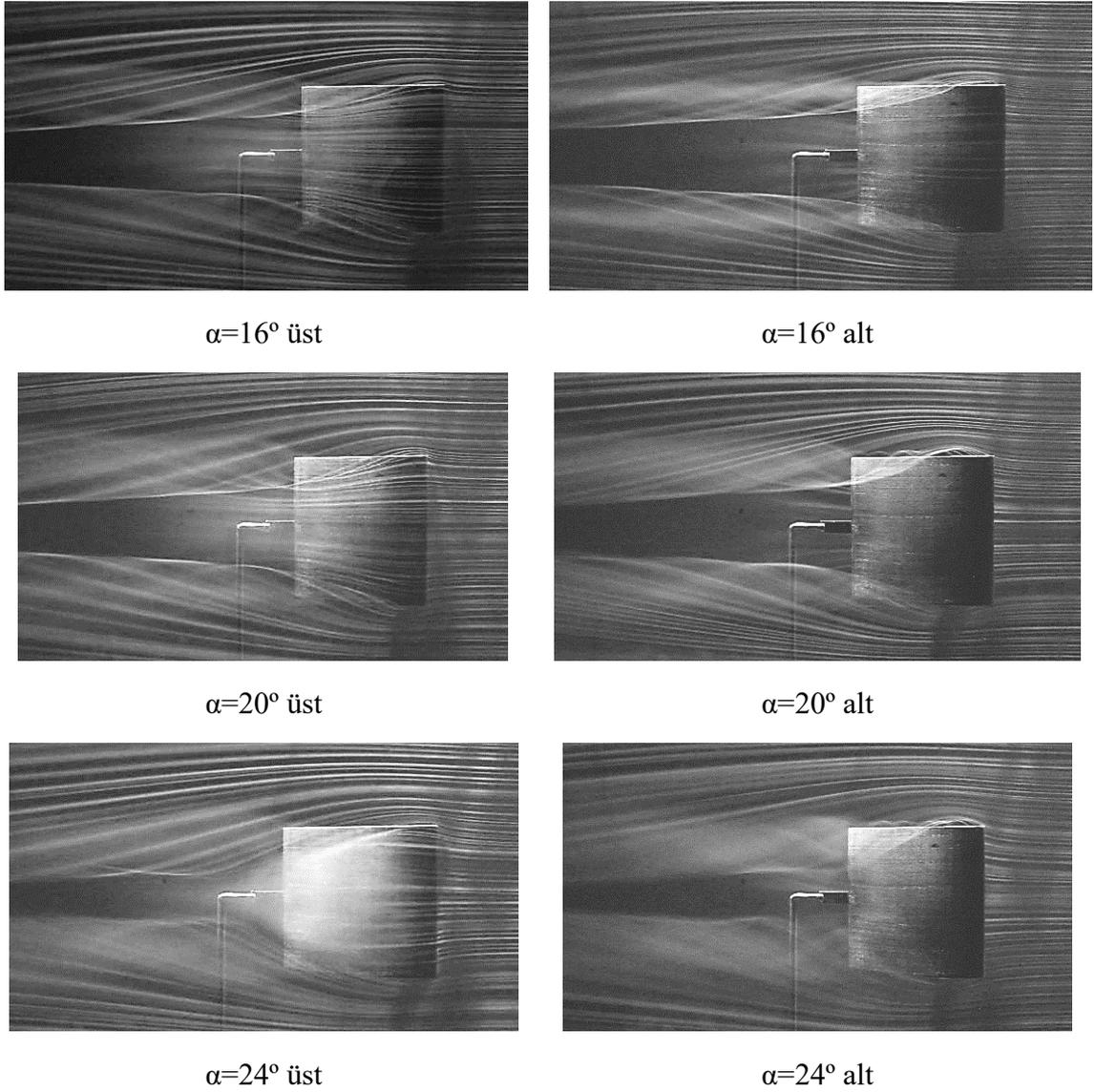

α=16º üst                    α=16º alt

α=20º üst                    α=20º alt

α=24º üst                    α=24º alt

Şekil 3.29. Duman teli ile akış görselleştirme Re=50000 (Devamı)

Duman teli deneyinde akış hızının artması ile yağ damlacıklarının tele tutunmaları zorlaşmış. Deneylerin yapılması sırasında tele damlatılan yağ damlacıkları akışın hızından dolayı telden koparak kanat üzerine damlamış ve kanadın yüzeyinde bir katman oluşturmuştur. Bu durum hem kanadın yapısında değişme meydana getirdiği hem de yeterli duman elde edilemediği için Re = 75000 için deneyler gerçekleştirilememiştir.

Genel sonuç çıkardığımızda; özellikle düşük Re sayılarında uç girdaplar kanat üzerindeki akışa ve ayrılma kabarcığına etki ederek kanadın aniden tutunma kaybına uğramasını engellemekte ve Re sayısı artığında ise uç girdabı büyüdüğü için olumlu etki olumsuza dönüşmekte ve aniden tutunma kaybına sebep olmaktadır.

### 3.2.8. Yağ Deneyi İle Akış Görselleştirme

Yapılan çalışmada akış görselleştirme tekniklerinden olan yağ ile akış görselleştirme tekniği kullanılmıştır. Gazyağı karışımında renklendirici olarak titanyum dioksit ve karışımın homojenliğini sağlamak için ise oleik asit kullanılmıştır. Oluşturulan bu karışım Şekil 3.30'da gösterildiği gibi kanat yüzeyine fırça yardımı ile sürülmüştür.

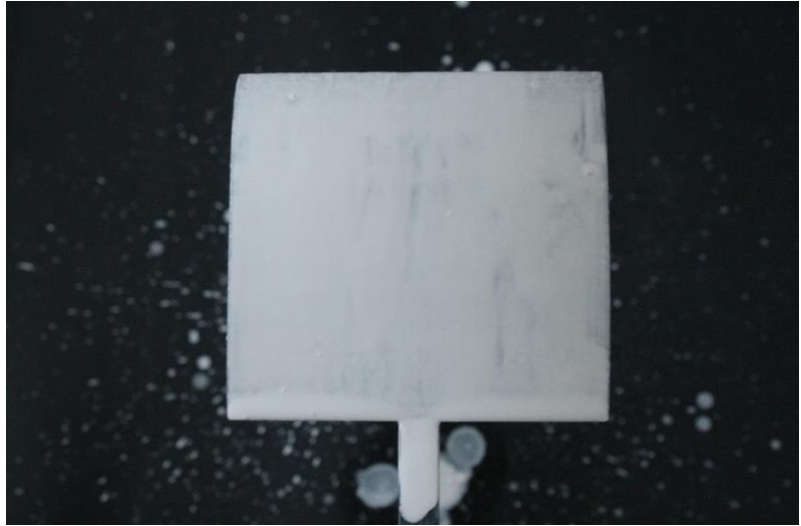

Şekil 3.30. Yağ deneyi için hazırlanmış NACA4412 profile sahip kanat

Bu işlem sırasında dikkat edilmesi gereken husus titanyum dioksit maddesinin kanat yüzeyinde pürüzleşmeye neden olmamasını sağlamaktır. Bu deneyde akış görselleştirme işlemi karışımın kanat üzerinden geçen hava akışı yardımı ile hareketinin sağlanmasına dayanmaktadır. Yöntemin çok yüksek maliyetli olmaması ve kolay şekilde gerçekleştirilmesine karşın düşük hızlarda karışımın viskozitesi yeterli olmadığı için deney yapılamamaktadır.

Şekil 3.31 ve 3.32'de görüldüğü gibi uç girdaplarının kanat üst kısmındaki etkisi Re=50000 ve 75000 için yağ deneyi ile görselleştirilmiştir. Akışın ayrılma ve tutunma noktaları ayrıca uç girdaplarının kanat yüzeyine etkisi bu deney yardımıyla görselleştirilmiştir. Ayrılma kabarcığı bölgesinde yağ karışımı hareketsiz olarak kalmış diğer bölgelerde yağ akışla birlikte kanat üzerinden akmıştır. Hücum açısının artırılması ile akış ayrılması hücum kenarına doğru ilerlemektedir. Bu durum benzer şekilde duman deneyi ile de örtüşmektedir. Hücum açısının artması uç girdapların kanat üzerindeki etkisini artırarak tutunma kaybının sonsuz kanat profilinde Re=50000 için AoA=16°'de [30] olması beklenirken yağ deneyinde AoA=24°'ye kadar akış ayrılması görülmemiştir. Benzer şekilde Re=75000 için akış ayrılması sonsuz kanatta AoA=18°'de [30] olmasına karşın AoA=24°'ye kadar bu ayrılma görülmemiştir.

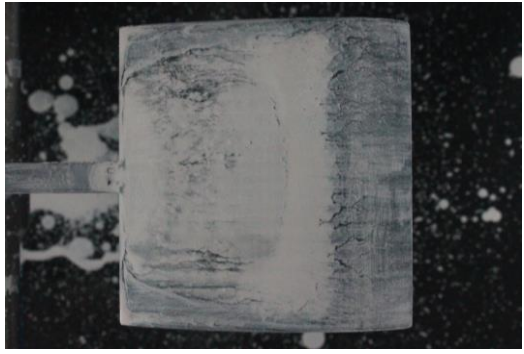
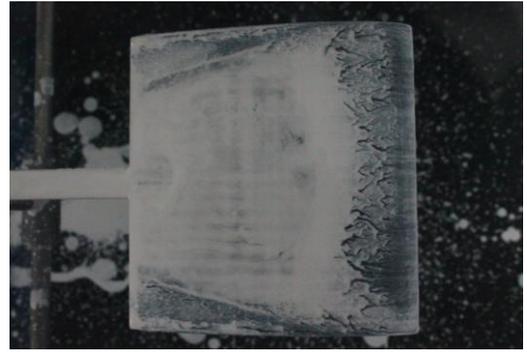

α=0°                    α=4°

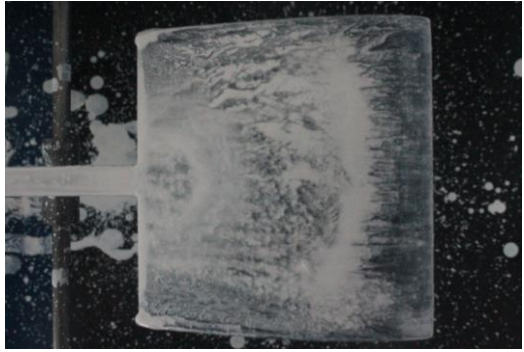
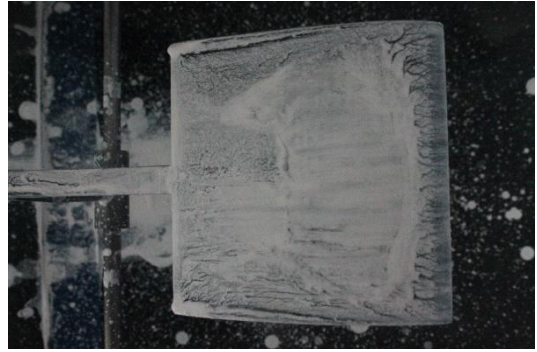

α=8°                    α=12°

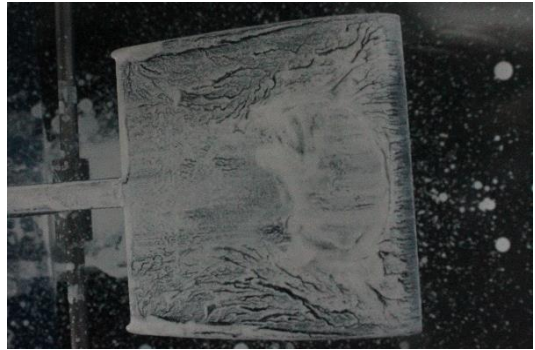
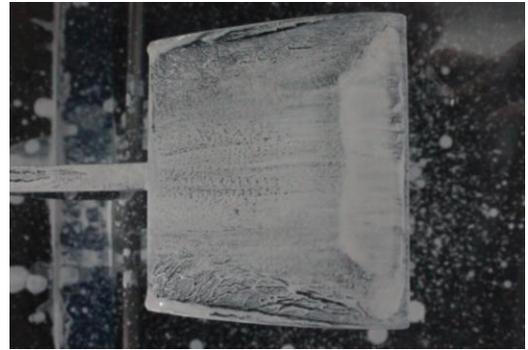

α=16°                   α=20°

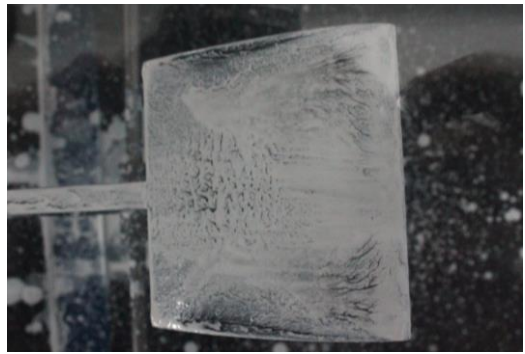

α=24°

Şekil 3.31. Re=50000 ve farklı açılarda yağ deneyleri

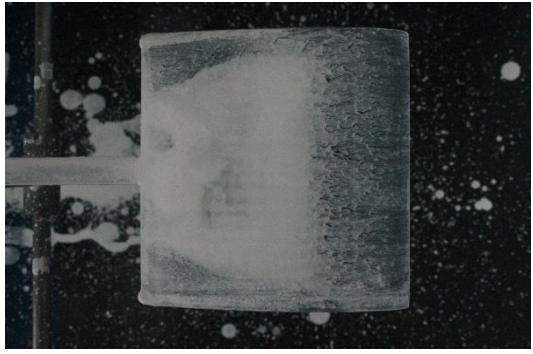
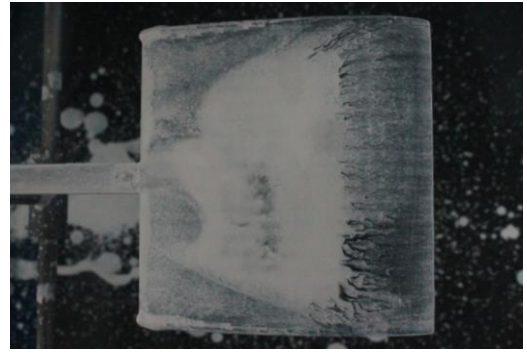

α=0°                              α=4°

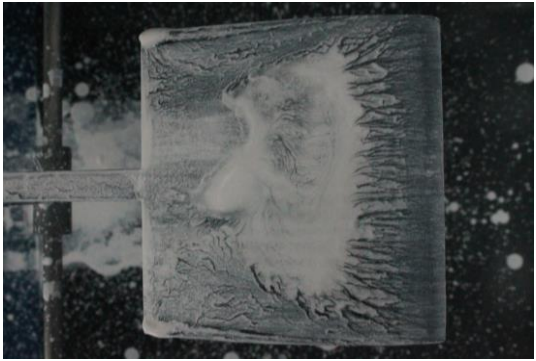
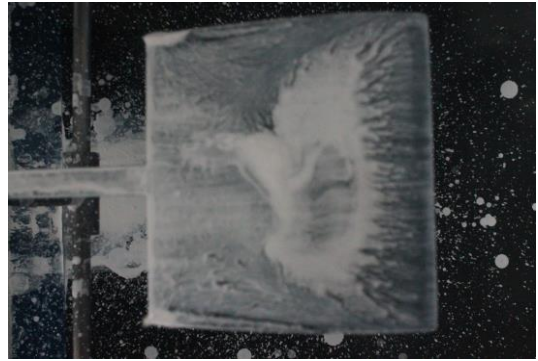

α=8°                              α=12°

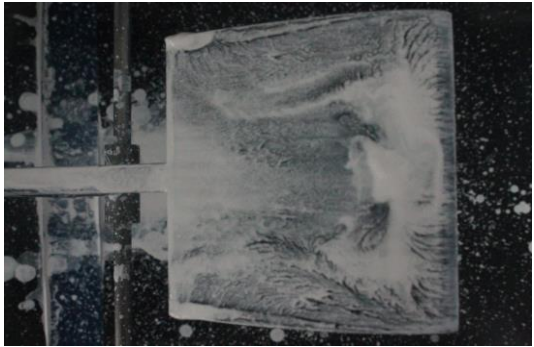
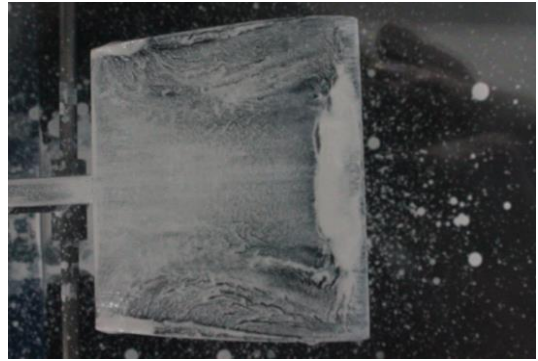

α=16°                             α=20°

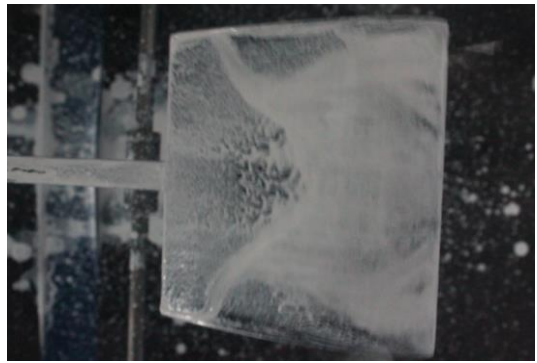

α=24°

Şekil 3.32. Re=75000 ve farklı açılarda yağ deneyleri

# 4. BÖLÜM

## TARTIŞMA – SONUÇ ve ÖNERİLER

### 4.1. Giriş

İnsansız ve mikro hava araçları, rüzgar türbinleri ve uçan canlıların düşük Re sayılarında çalışmasından dolayı bu konuda bir çok araştırmalar yapılmakta ve yapılan bu çalışmalar yardımı ile literatüre yeni bilgileri katılmaktadır. Askeri alanda insansız ve mikro hava araçlarının günümüz ve geleceğin teknolojisi olması ve yenilenebilir enerji kaynaklarına duyulan ihtiyacın her geçen gün artması bu alanda yapılan araştırmalara ihtiyacı artırmıştır.

Düşük Re sayılarında çalışan mikro hava araçlarının özellikle boyutlarının küçük olması düşük en-boy oranına sahip kanatların kullanılması zorunluluğunu da beraberinde getirmiştir. Bu yüzden düşük en-boy oranına sahip kanadın kararsız aerodinamiği yapılan bu tez çalışmasında incelenmiş özellikle uç girdaplarının olumlu ve olumsuz etkileri araştırılmıştır.

### 4.2. Sonuçlar

Uç girdapları, basınç gradyenleri nedeni ile kanadın alt yüzeyinde meydana gelen yüksek basınç kanadın üst yüzeyinde bulunan düşük basınç bölgesine doğru hareket etmeye başlar. Sonsuz kanat profilleri üzerinde görülen akışlarda görülen iki yönlü akış profilleri (x ve y yönünde), en-boy oranına sahip kanatlarda uç girdapları ektisi ile üç boyutlu (x, y ve z yönünde) yapıya sahip olur. En-boy oranı küçüldükçe uç girdapların etkisi kanat yüzeyine yayıldığı için bu etki daha fazla olur. Bu girdaplar kanat üzerinde meydana gelen kaldırma ve sürükleme kuvvetlerini artırdığı deneysel çalışmalardan da görülmüştür. Literatür araştırmasında Genç ve arkadaşlarının [30] yapmış olduğu çalışmada sonsuz NACA4412 kanat profilinin deneysel çalışması yapılmış ve kaldırma kuvvet katsayıları Re=25000, 50000 ve 75000 için sırasıyla 0.9, 1.25, ve 1.35 olarak ve tutunma kaybı açıları 12°, 16° ve 18° [30] bulunmuştur. Yaptığımız çalışma sonucunda kaldırma kuvvet sayıları 1.17, 1.31 ve 1.32 olarak elde edilmiştir. Ayrıca tutunma kaybı açılarını sırasıyla 39°, 38° ve 37°'ye gecikmiştir. Bu çalışmadan da anlaşıldığı üzere uç girdaplar kanat üzerindeki ayrılmalara ve ayrılma kabarcığına etkisi görülmüştür.

## 4.3.    Öneriler

Yapılan bu çalışmada kanat üzerindeki hız dağılımı için sıcak tel anemometresi kullanılmıştır. Bu deneylerin çok zaman alması ve deney sırasındaki bazı aksaklıkların (probun kırılması, kalibrasyonun sürekli bozulması ve probun kirlenmesi) meydana gelmesinden dolayı sıcak tel anemometresi yerine PIV sistemi kullanılabilir.

Yapılan deneysel çalışmanın sayısal çözümlemeleri yapılarak uygun yöntem belirlenebilir ve sayısal tutarlılık incelenebilir.

Yapılan bu çalışmada sadece AR=1 NACA4412 profiline sahip kanat kullanılmıştır. Bu yüzden daha düşük veya büyük en-boy oranına sahip kanatlar kullanılabilir. Böylelikle en-boy oranının akış üzerindeki etkisi incelenebilir ve en uygun en-boy oranları belirlenebilir. Ayrıca bilindiği gibi NACA4412 kambur bir kanattır. Bu yüzden simetrik bir kanat ile bu çalışma yapılarak kamburluğun uç girdaplara ve aerodinamik performansa etkisi incelenebilir.

# KAYNAKLAR

[1]    Genç, M. S., Karasu, I., Açikel, H. H., 2012. An experimental study on aerodynamics of NACA2415 aerofoil at low Re numbers, **Experimental Thermal and Fluid Science**, **39**: 252–264.

[2]    Katz, J., Galdo, J. B., 1989. Effect of roughness on rollup of tip vortices on a rectangular hydrofoil. **Journal of Aircraft, 26** (3): 247–253.

[3]    Viieru, D., Lian, Y., Shyy, W., Ifju, P., 2003. Investigation of tip vortex on aerodynamic performance of a micro air vehicle. **AIAA paper, 3597**: 1-10.

[4]    Gim, O. S., Lee, G. W., 2013. Flow characteristics and tip vortex formation around a NACA 0018 foil with an endplate. **Ocean Engineering, 60**: 28–40.

[5]    Liu, Y.,  Hsiao, F., 2013. Experimental Investigation on Critical Reynolds Numbers Aerodynamic Properties of Low Aspect Ratios Wings. **Procedia Engineering, 00(1st ICM)**: 1–10.

[6]    Alam, M.,  Sandham, N., 2000. Direct numerical simulation of "short"laminar separation bubbles with turbulent reattachment. **Journal of Fluid Mechanics, 403**: 223–250.

[7]    Freymuth, P., Bank, W., Finaish, F., 1986. Visualization of wing tip vortices in accelerating and steady flow. **Journal of Aircraft, 23** (9): 730–733.

[8]    Giuni, M., Green, R. B., 2013. Vortex formation on squared and rounded tip. **Aerospace Science and Technology, 29** (1): 191–199.

[9]    Vaidya, H. A., Kevadiya, M., 2013. 2D Analysis of Naca 4412 Airfoil, **International Journal of Innovative Research in Science, Engineering and Technology 2** (5): 1686–1691.

[10]   Mueller, T. J., 1999. Aerodynamic Measurements at Low Reynolds Numbers for Fixed Wing MAVs. **Development and Operations of UAVs for Military and Civil Applications**, 1–32.

[11]   Birch, D., Lee, T., Mokhtarian, F., Kafyeke, F., 2004. Structure and Induced Drag of a Tip Vortex. **Journal of Aircraft, 41** (5): 1138–1145.

[12]   Ramaprian, B. R., Zheng, Y., 1997. Measurements in Rollup Region of the Tip Vortex from a Rectangular Wing. **AIAA Journal, 45** (12): 1837–1843.

[13]   Liang, Z., Xue, L., 2014. Detached-eddy simulation of wing-tip vortex in the near field of NACA 0015 airfoil. **Journal of Hydrodynamics, Ser. B, 26** (2): 199–206.

[14]   Yarusevych, S.,  Boutilier, M. S. H., 2010. Vortex Shedding Characteristics of a NACA 0018 Airfoil at Low Reynolds Numbers. pp. 1–8. *40th Fluid Dynamics Conference and Exhibit*, June 28 - July 01, 2010 Chicago, Illinois.

[15]   Mueller, T. J., Delaurier, J. D., 2003. Aerodynamics of Small Vehicles, **Annual Review of Fluid Mechanics, 35**: 89–111.

[16]   Ho, S., Nassef, H., Pornsinsirirak, N., Tai, Y. C., Ho, C. M., 2003. Unsteady aerodynamics and flow control for flapping wing flyers. **Progress in Aerospace Sciences, 39**: 635–681.

[17]   Woods, M. I., Henderson, J. F., Lock, G. D., 2001. Energy requirements for the flight of micro air vehicles. **The Aeronautical Journal, 105** (1045): 135–149.

[18]   Gursul, I., 2004. Vortex flows on UAVs: Issues and challenges. **Aeronautical Journal, 108** (1090): 597–610.

[19]   Whitehead, J., Gursul, I., 2006. Interaction of Synthetic Jet Propulsion with Airfoil Aerodynamics at Low Reynolds Numbers, **AIAA Journal, 44** (8), 1753–1766.

[20]   Rojratsirikul, P., Wang, Z., Gursul, I., 2008. Unsteady Aerodynamics of Membrane Airfoils, pp. 7–10. *In 46th AIAA Aerospace Sciences Meeting and Exhibit*, January 7 - 10, Reno, NV, USA.

[21]   Rojratsirikul, P., Wang, Z., Gursul, I., 2009. Unsteady fluid – structure interactions of membrane airfoils at low Reynolds numbers, **Experiments in Fluids, 46** (5): 859–872.

[22]   Kaplan, S. M., Altman, A., Ol, M., 2007. Wake Vorticity Measurements for Low Aspect Ratio Wings at Low Reynolds Number. **J Aircraft, 44** (1): 241–251.

[23]   Torres, G. E., Mueller, T. J., 2004. Low-Aspect-Ratio Wing Aerodynamics at Low Reynolds Numbers. **AIAA Journal, 42** (5): 865–873.

[24]   Pelletier, A., Mueller, T. J., 2000. Low Reynolds Number Aerodynamics of Low-Aspect-Ratio , Thin / Flat / Cambered-Plate Wings Introduction. **Journal of**

**Aircraft, 37** (5): 825–832.

[25] Genç, M.S., Özışık, G., Kahraman, N., 2008 Düz Flaplı NACA0012 Kanat Profilinin Aerodinamik Performansının İncelenmesi, **Isı Bilimi ve Teknigi Dergisi-J. Therm. Sci. Technol, 28** (1): 1–8.

[26] Özkan, G., 2014. Rüzgar Tüneli İçin Kuvvet Ölçüm Sistemi ve Açı Mekanizması Tasarımı. Erciyes Üniversitesi, Mühendsilik Fakültesi, Bitirme Ödevi, Kayseri, 33 s.

[27] Yükselen, M. A., 2006. Aerodinamik kuvvet ve momentler. (Web sayfası: **http://web.itu.edu.tr/~yukselen/Uck351/02-%20Aerodinamik%20kuvvet%20ve %20momentler.pdf** (Erişim tarihi: Mayıs 2015)

[28] Karasu, İ., 2011. Düşük Reynolds Sayılı Akışlarda Kanat Profili Üzerinde Türbülansa Geçişin Ve Laminer Ayrılma Kabarcığının Deneysel Ve Sayısal İncelenmesi. Erciyes Üniversitesi, Fen Bilimleri Enstitüsü, Yüksek Lisans Tezi, Kayseri, 89 s.

[29] Williamson, C. H. K., 1996. Vortex Dynamics in the Cylinder Wake. **Annual Review of Fluid Mechanics, 28** (1): 477–539.

[30] Genç, M. S., Koca, K., Açıkel, H. H., Özkan, G., Kırış, M. S., Yıldız, R., 2016. Flow Characteristics over NACA4412 airfoil at Low Reynolds Number. *10th Experimental fluid mechanics,* November 17-20, 2015, Prague, Czech Republic.

# ÖZGEÇMİŞ

**KİŞİSEL BİLGİLER**

**Adı, Soyadı:** Gökhan ÖZKAN
**Uyruğu:** Türkiye (TC)
**Doğum Tarihi ve Yeri:** 1 Aralık 1984, Kayseri
**Medeni Durumu:** Bekar
**Tel:** +90 352 224 88 00
**Fax:** +90 352 224 88 00
**email:** gokhan.ozkan@agu.edu.tr
**Yazışma Adresi:** Abdullah Gül Üniversitesi Sümer Kampüsü, Mühendislik Fakültesi Elektrik ve Bilgisayar Mühendisliği Bölümü, 38010 KAYSERİ

**EĞİTİM**

| Derece | Kurum | Mezuniyet Tarihi |
|--------|-------|------------------|
| Lisans | ERÜ Enerji Sistemleri Mühendisliği | 2014 |
| Lisans | Marmara Üni. Elektrik Öğretmneliği | 2006 |
| Lise | Anadolu Meslek Lisesi, Nevşehir | 2002 |

**İŞ DENEYİMLERİ**

| Yıl | Kurum | Görev |
|-----|-------|-------|
| 2014–2015 | Bozok Üniversitesi Akdağmadeni Meslek Yüksek Okulu Elektrik ve Enerji Bölümü | Öğretim Görevlisi |
| 2012–2015 | Erciyes Üniversitesi Mühendislik Fakültesi Enerji Sistemleri Mühendisliği Bölümü | Proje Asistanı |
| 2010–2011 | Atatürk Ticaret Meslek Lisesi, Kayseri | Branş Öğretmeni |

**BAŞARI ve ÖDÜLLER**

| 2014 | Erciyes Üniversitesi Mühendislik Fakültesi Fakülte Birinciliği |
|------|------|
| 2014 | Erciyes Üniversitesi Enerji Sistemleri Mühendisliği Bölüm Birinciliği |
| 2012-2014 (5 Kez) | Erciyes Üniversitesi Enerji Sistemleri Mühendisliği Yüksek Onur Öğrencisi |

**YABANCI DİL**
İngilizce